%% file: main.tex
\documentclass[aps,prb,twocolumn,english,showpacs,letterpaper,superscriptaddress,citeautoscript]{revtex4-1}

\usepackage{lineno}

\usepackage{amsmath}
\usepackage{amssymb}
\usepackage[dvips]{epsfig}
\usepackage{graphicx}
\usepackage{subfigure}
\usepackage{array}
\usepackage{physics}

\usepackage{color}
\usepackage[colorlinks=true,citecolor=blue,linkcolor=blue]{hyperref}
\usepackage{float}
\usepackage{mathrsfs}
\usepackage{indentfirst}
\usepackage{textcomp}
\usepackage{comment}
\usepackage{mathtools}

\usepackage{verbatim}

\newcommand{\ignore}[1]{}

\newcommand{\tpm}{t^\prime}
\newcommand{\tppm}{t^{\prime\prime}}
\newcommand{\tp}{$t^\prime$}
\newcommand{\tpp}{$t^{\prime\prime}$}

\newcommand{\tttJ}{$t$-{\tp}-{\tpp}-$J$}
\newcommand{\kbf}{\mathbf{k}} 
\newcommand{\qbf}{\mathbf{q}}

\begin{document}


\title{Machine-learning the spectral function of a hole in a quantum antiferromagnet}
\author{Jackson Lee}
\affiliation{Physics and Computer Science Department, Rutgers University, New Brunswick, New Jersey 08854, USA}
\affiliation{Condensed Matter Physics and Materials Science Division, Brookhaven National Laboratory, Upton, New York 11973, USA}
\author{Matthew R. Carbone}
\affiliation{Computational Science Initiative, Brookhaven National Laboratory, Upton, New York 11973, USA}
\author{Weiguo Yin}
\email{wyin@bnl.gov}
\affiliation{Condensed Matter Physics and Materials Science Division, Brookhaven National Laboratory, Upton, New York 11973, USA}

\date{\today}

\begin{abstract}
Understanding charge motion in a background of interacting quantum spins is a fundamental problem in quantum many-body physics. The most extensively studied model for this problem is the so-called {\tttJ} model, where the determination of the parameter {\tp} in the context of cuprate superconductors is challenging. Here we present a theoretical study of the spectral functions of a mobile hole in the {\tttJ} model using two machine learning techniques: K-nearest Neighbors regression (KNN) and a feed-forward neural network (FFNN). We employ the self-consistent Born approximation to generate a dataset of about $1.3 \times 10^5$ spectral functions. We show that for the forward problem, both methods allow for the accurate and efficient prediction of spectral functions, allowing for e.g. rapid searches through parameter space. Furthermore, we find that for the inverse problem (inferring Hamiltonian parameters from spectra), the FFNN can, but the KNN cannot, accurately predict the model parameters using merely the density-of-state. Our results suggest that it may be possible to use deep learning methods to predict materials parameters from experimentally measured spectral functions.

\end{abstract}

\maketitle
\section{Introduction}

\begin{figure*}[t]
    \begin{center}
            \includegraphics[width=0.95\textwidth]{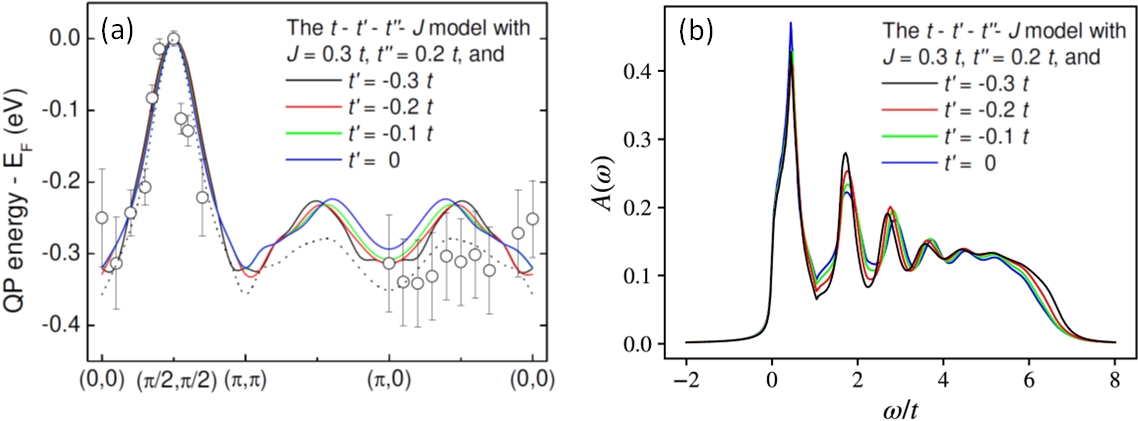}
    \end{center}
    \caption{The $t'$ dependence of (a) the quasiparticle band dispersion $E(\mathbf{k})$ compared with experimental data (open circles), reproducing Fig.~4 in Ref.~\onlinecite{Yin_PRB_09_WF} with permission, and (b) the density of states $A(\omega)=\sum_\mathbf{k} A(\mathbf{k},\omega)$, where the first broad peaks around the Fermi level (zero energy) are almost identical for a wide range of $t'$ but the other peaks could be used to resolve $t'$. 
    }
    \label{fig:problem}
\end{figure*}

Understanding charge motion in a background of interacting quantum spins has been considered an essential first step in search for the mechanisms of superconductivity in many unconventional materials ranging from cuprates~\cite{Dagotto_RMP_94,Lee_RMP_06,Schmitt-Rink_PRL_88_SCBA} to iron-based superconductors~\cite{Yin_book_15,Yin_PRL_10_FeTe} and to twisted bilayer graphene~\cite{Cao_Nature_18_TBG_SC}. This topic has recently received renewed interest thanks to recent novel experiments using ultracold atoms in optical lattices, as they provide an essentially perfect realization of the Fermi-Hubbard model, with site-resolved imaging ability~\cite{Ji_PRX_21,Koepsell_Science_21_cold-atom,Koepsell_Nature_19_cold-atom,Bohrdt_NP_19_ML_snapshots,Chiu_Science_19_cold-atom,Brown_Science_19_cold-atom,Mazurenko_Nature_17_cold-atom,Nyhegn_PRB_22_bilayer-t-J_SCBA_dynamics}. The most extensively studied model for this problem is the so-called $t$-$J$-type model~\cite{Zhang-Rice_PRB_88}. Its applicability to cuprates was established by comparing the model study results with various experiments, most notably angle-resolved photoemission spectroscopy (ARPES), which can specifically yield the spectral function of holes introduced by photoemission of electrons~\cite{Damascelli_RMP_03_reivew_arpes,Marshall_PRL_96,Eder_PRB_97,Kim_PRL_98,Yin_PRL_98}. The single-hole problem corresponds to photoemission from an undoped Mott insulator, such as Sr$_2$CuO$_2$Cl$_2$ (a parent compound of cuprates), where besides the nearest-neighbor hole hopping parameter ($t$), the second and third nearest-neighbor hopping parameters ({\tp} and \tpp) are found to be necessary to reproduce the correct quasiparticle dispersion relation $E(\mathbf{k})$~\cite{Wells_PRL_95_Sr2CuO2Cl2,Ronning_PRB_03_undoped_arpes,Nazarenko_PRB_95_Sr2CuO2Cl2,Kyung_PRB_96_Sr2CuO2Cl2,Xiang_PRB_96_Sr2CuO2Cl2,Yin_PRB_09_WF,Belinicher_PRB_96_Sr2CuO2Cl2,Leung_PRB_97_Sr2CuO2Cl2,Lee_PRB_97_Sr2CuO2Cl2}. While the need for longer hopping parameters is justified by first-principles analysis of the in-crystal overlapping of electronic wave functions~\cite{Yin_PRB_09_WF,Pavarini_PRL_01_WF}, it was uncovered~\cite{Yin_PRB_09_WF,Belinicher_PRB_96_Sr2CuO2Cl2} that the determination of {\tp} via fitting $E(\mathbf{k})$ is inconclusive because $E(\mathbf{k})$ could be insensitive to {\tp} varying from $0$ to $-0.3t$ [see Fig.~\ref{fig:problem}(a)].
This is a relevant problem since the value of {\tp} was shown to correlate with the superconducting transition temperature at optimal doping~\cite{Pavarini_PRL_01_WF} and affect phase competition~\cite{Yin_PRB_17_phase-competition} in the cuprates.

Another drawback of this traditional approach to predicting model parameters is that $E(\mathbf{k})$ is generally derived from low-energy spectral peaks. However, it is difficult to resolve $E(\mathbf{k})$ when the quasiparticle spectral weight is small~\cite{Valla_Science_99_MDC,Laughlin_PRL_97_Sr2CuO2Cl2_slave-boson}, which is a common phenomenon in systems close to a non-Fermi liquid state, specifically near the high-energy edge of the quasiparticle band in cuprates, resulting in large error bars [see Fig.~\ref{fig:problem}(a)]. It is thus highly desirable if the model parameters can be predicted by studying the full energy range of the spectral functions directly [see Fig.~\ref{fig:problem}(b)]. Extension from fitting $E(\mathbf{k})$ to treating the whole spectral function $A(\mathbf{k},\omega)$ means a dramatic increase in the total amount of data that needs to be processed. Here we use machine learning (ML) methods to address this outstanding problem in the field of strongly correlated electron systems and high-temperature superconductivity.

Machine learning plays a huge role when cataloguing or processing large amounts of data in general. It is uniquely able to identify important patterns and correlations that might otherwise be missed, especially in large datasets. Recently, it has emerged as an important computational tool across disciplines in the physical sciences~\cite{Krenn_NRP_22_ML_review}. For example, in particle physics, ML played an instrumental role in the discovery of the Higgs boson
~\cite{Radovic_Nature_18_ML_ParticlePhys}. In astrophysics, ML techniques have been used to study photometric redshifts
~\cite{Sadeh_16_ML_redshift}, cluster membership of galaxies~\cite{Hashimoto_Universe_22_ML_galaxy}, and exoplanet transit detection
~\cite{Schanche_19_ML_ExoplanetTransit}. In materials and molecular science, ML is heralding in a ``second computational revolution''~\cite{Schmidt_npjCM_19_ML_review}, helping predict crystal structures~\cite{Ryan_18_JACS_ML_crystalStruct}, calculate material properties~\cite{Zheng_ChemSci_18_ML_properties,carbone2019classification,torrisi2020random}, and accelerate first-principles calculations~\cite{Jalem_SR_18_ML_fastIonConduct,carbone2020machine,ghose2022uncertainty,rankine2022accurate,penfold2022deep}. In condensed matter physics, ML was used to find phase transition temperatures~\cite{Carrasquilla_NP_17_ML_phases}, catalogue snapshots of strongly correlated electronic states~\cite{Bohrdt_NP_19_ML_snapshots}, infer fundamental physical information from model systems~\cite{Miles_PRB_21_ML_impurity}, efficiently sample configurations in many-body systems~\cite{Liu_PRB_17_ML_MonteCarlo}, and predict impurity spectral functions~\cite{Sturm_PRB_21_ML_impurity}. 

In this paper, we show how ML can be used to predict and understand spectral functions in the {\tttJ} model. The paper is organized as follows: Section~\ref{Methods} describes the {\tttJ} model, the used ML methods, and how we obtain the needed dataset for training, validation, and testing. Section~\ref{PCA} presents a preliminary examination of the data using Principal Component Analysis (PCA), primarily to help determine what linear correlations may be present in the data. Section~\ref{forward} addresses the \emph{forward} problem of predicting spectral functions from a given set of model parameters~\cite{Sturm_PRB_21_ML_impurity,Arsenault_PRB_14_ML_impurity,walker2020neural}. Section~\ref{inverse} addresses the \emph{inverse} problem of predicting the model parameters $t'/t$, $t''/t$, and $J/t$ from spectral functions. Section~\ref{inverse_t} introduces an algorithm to find the value of $t$. Finally, the results and discussions are summarized in Section~\ref{Summary}.

\section{Methods\label{Methods}}

\subsection{Hamiltonian and spectral functions}

The {\tttJ} model is described by the following Hamiltonian~\cite{Yin_PRL_98}:
\begin{eqnarray}
H = &-& \left(t\sum_{\langle i, j \rangle_1, \sigma} + t'\sum_{\langle i, j \rangle_2, \sigma} + t''\sum_{\langle i, j \rangle_3, \sigma}\right)\left(\tilde{c}^{\dagger}_{i\sigma} \tilde{c}^{}_{j,\sigma}+h.c.\right) \nonumber \\
& + & J\sum_{\langle i,j \rangle_1}{\textbf{S}_i \cdot \textbf{S}_j} 
\end{eqnarray}
in the standard notation of the constrained fermionic operators: $\tilde{c}^{\dagger}_{i\sigma}$ creates an electron with the spin index $\sigma$ (either $\uparrow$ or $\downarrow$) at site $i$---with the constraint of no double occupancy at any site---and $\tilde{c}^{}_{i,\sigma}$ annihilates it. The spin operators $\mathbf{S}_i$ expressed in the matrix form are given by $\left(\mathbf{S}_i\right)_{\sigma \sigma^\prime}=\frac{1}{2}\sum_{\sigma \sigma^\prime}{\tilde{c}^{\dagger}_{i\sigma} \hat{\tau}^{}_{\sigma \sigma^\prime} \tilde{c}^{}_{i\sigma^\prime}}$ where $\hat{\tau}=(\hat{\tau}^x,\hat{\tau}^y,\hat{\tau}^z)$ are the $2\times2$ Pauli matrices. 
The angle brackets denote the first ($\langle i,j \rangle_1$), second ($\langle i,j \rangle_2$), and third ($\langle i,j \rangle_3$) neighbor sites, respectively.
Thus, the $J$ term describes the Heisenberg interaction between nearest-neighboring quantum spins; the $t$, \tp, and {\tpp} terms describe the electron hopping to nearest, second nearest, and third nearest sites, respectively. 

The angle-resolved spectral function of a doped hole with momentum $\mathbf{k}$ and energy $\omega$ is given by
\begin{equation}
A(\mathbf{k}, \omega) = -\frac{1}{\pi} \Im G(\mathbf{k}, \omega),
\label{eq:A}
\end{equation}
with the retarded Green's function of the single hole being
\begin{equation}
 G(\mathbf{k}, \omega) =  \lim_{\eta \to 0^+} \langle \Psi_0 | \tilde{c}^{\dagger}_{\mathbf{k}\sigma} \frac{1}{\omega + i\eta - H + E_0} \tilde{c}^{}_{\mathbf{k}\sigma}|\Psi_0 \rangle,
 \label{eq:G}
\end{equation}
where $E_0$ and $|\Psi_0 \rangle$ are the ground-state energy and wave function of the undoped system, respectively, thus $H|\Psi_0 \rangle=E_0|\Psi_0 \rangle$. Or equivalently, 
\begin{equation}
A(\mathbf{k}, \omega) = \sum_\nu {|\langle \nu | \tilde{c}^{}_{\mathbf{k}\sigma}|\Psi_0 \rangle|^2 \delta(\omega - E_\nu + E_0) },
\label{ARPES}
\end{equation}
where $|\nu \rangle$ is an eigenstate of $H$ with one less electron and $E_\nu$ is the corresponding eigen-energy satisfying $H|\nu \rangle=E_\nu |\nu \rangle$, as the Dirac delta function $\delta(\omega)$ is related to a Lorentzian by $\delta(\omega)=\lim_{\eta \to 0^+} \frac{1}{\pi} \frac{\eta}{\omega^2+\eta^2}=\lim_{\eta \to 0^+} -\frac{1}{\pi} \mathrm{Im}\,\frac{1}{\omega+i\eta}$. 

The angle-\emph{unresolved} spectral function is given by
\begin{equation}
A(\omega) = \sum_{\mathbf{k}}A(\mathbf{k}, \omega),
\label{DOS}
\end{equation}
which is also called the density of states (DOS). To obtain the DOS in the normal procedure of theoretical calculations, one needs to have first calculated out $A(\mathbf{k}, \omega)$ using Eq.~(\ref{ARPES}) for a dense mesh of $\mathbf{k}$ points, and then sum the results over $\mathbf{k}$ using Eq.~(\ref{DOS}). This implies that if DOS can be accurately predicted from known DOS data, a significant speedup in evaluating DOS can be achieved, e.g., a four-orders-of-magnitude speedup compared with the normal procedure using a $100\times 100$ $\mathbf{k}$-mesh. More interestingly, we will explore whether the model Hamiltonian parameters can be accurately predicted by machine learning the DOS $A(\omega)$, which is relevant to x-ray photoemission (XPS), or $A(\mathbf{k}, \omega)$ with a fixed $\mathbf{k}$, which is relevant to laser-based ARPES where the $\mathbf{k}$ points are most accessible near the zone center $\mathbf{k}=0$.

\subsection{Dataset generation}

To obtain the dataset for use in our ML approach, we use the self-consistent born approximation (SCBA) to calculate Green's function of a hole in the {\tttJ} model~\cite{Schmitt-Rink_PRL_88_SCBA,Marsiglio_PRB_91_SCBA,Martinez_PRB_91_SCBA,Liu_PRB_91_SCBA,Liu_PRB_92_SCBA,Yin_PRB_97_SCBA_bilayer,Yin_PRB_98_SCBA_multilayer,Manousakis_PRB_07_kink_SCBA,Manousakis_PLA_07_kink_SCBA,Valla_PRL_07_kink_SCBA} (see Appendix~\ref{appendix_scba} for details). This approximation produces quantitatively accurate results for the hole Green’s function compared with exact diagonalization on
small systems~\cite{Leung_PRB_95_ED_SCBA} and Monte Carlo simulations~\cite{Diamantis_NJP_21_SCBA_MonteCarlo}.

We note that although the Hamiltonian has four parameters $(t,\tpm,\tppm, J)$, all the data can be scaled with respect to $t$, e.g. 
\begin{equation}
\label{eq:scale}
A(\mathbf{k},\omega) \to A(\mathbf{k},a\,\omega)/a \quad\mathrm{for}\quad t \to a\,t.
\end{equation}
where $a$ is an arbitrary positive real number.  Setting $t$ as the energy unit ($t=1$) reduces the ML complexity by one dimension, which is of significant advantage in high-throughput computation and big data management. Thus, the Green's functions are generated in a grid of $\tpm \in [-0.5, 0.5]$, $\tppm \in [-0.5, 0.5]$ and $J \in [0.2, 1.0]$, with each parameter sampled on a 51-point uniform grid. 

For each combination of \tp, \tpp, and $J$, the calculation of the Green's function $G(\mathbf{k}, \omega)$ is performed by using a $128\times 128$ mesh for the $\mathbf{k}$ points, $\omega \in [-6, 6]$ with the step (i.e., energy resolution) being 0.01, and $\eta=0.01$. Then, $\eta=0.1$ is used to broaden the resulting spiky DOS and a uniform grid of 301 $\omega$ points is used to sample the DOS. Therefore, our dataset for the DOS consists of $51^3=132,651$ pairs $\left(\textbf{x}^{(i)}, \textbf{y}^{(i)}\right)$, with $\mathbf{x}^{(i)} = (\tpm, \tppm, J)^{(i)}$ being the 3-dimensional vector representation of the $i$th model parameter set and $\mathbf{y}^{(i)}=(A(\omega_1),A(\omega_2),\dots,A(\omega_{301}))^{(i)}$ the corresponding 301-dimensional vector representation of the DOS. For the forward problem, $\mathbf{x}^{(i)}$ are the input and $\mathbf{y}^{(i)}$ are the output. For the inverse problem, the definitions of $\textbf{x}^{(i)}$ and $\mathbf{y}^{(i)}$ are switched, i.e.,  $\mathbf{x}^{(i)}=(A(\omega_1),A(\omega_2),\dots,A(\omega_{301}))^{(i)}$ and $\mathbf{y}^{(i)} = (\tpm, \tppm, J)^{(i)}$. We then randomly partitioned the dataset into an 80/10/10 training ($\mathbb{T}$), validation ($\mathbb{V}$), and testing $\mathcal{T}$ split. Here we use the computationally generated testing sets to demonstrate without any ambiguity that the ML methods work well for the present baseline problems.

\subsection{Machine Learning Methods}

Training a ML model consists of an optimization procedure in which a loss function encoding the difference between predicted and ground-truth outputs is minimized on a training set. In addition, a set of hyperparameters of the ML model is tuned during cross-validation to achieve high accuracy on the validation set. Hyperparameters are untrained parameters that include, but are not limited to, training time, network architecture and activation functions. Ultimately, final results are presented on the testing set in order to provide an unbiased estimate of model performance. Here we use the total mean squared error (MSE) as the loss function. Given the training set $\mathbb{T}=\left\{\left(\mathbf{x}^{(i)}, \mathbf{y}^{(i)}\right)\right\}$ of size $|\mathbb{T}|$, i.e.,  $i=1,2,3,...,|\mathbb{T}|$, for an $n$-dimensional input vector $\mathbf{x}^{(i)}$, the corresponding ground-truth output is a $m$-dimensional vector $\mathbf{y}^{(i)}$;  if the ML model predicts $\hat{\mathbf{y}}^{(i)}$, then the individual MSE for that training example is given by
\begin{equation}
    L^{(i)} = \frac{1}{m} \sum_{j=1}^m \abs{\hat{y}_j^{(i)}-y_j^{(i)}}^2,
\end{equation}
and the total MSE score of the ML method is given by
\begin{equation}
L = \frac{1}{\abs{\mathbb{T}}} \sum_{i=1}^{\abs{\mathbb{T}}} L^{(i)}.
\end{equation}


We now introduce the two ML methods used in this work: K-nearest neighbors (KNN) and feed-forward neural network (FFNN). 

\subsubsection{K-nearest neighbors}

\ignore{
\begin{figure}
\centering
\includegraphics[width=1.0\textwidth]{knn visualization.PNG}
\caption{\label{fig:knn visualization} Visualization of the k-nearest neighbors algorithm in a two-dimensional input parameter space for $k=4$. The blue points represent the input data points with known output values, and the orange point an input data point with unknown output. The 4 blue points nearest to the orange point are the ones used to determine the output value of the orange point.}
\end{figure}
}

The KNN algorithm predicts an output via a nearest neighbor search~\cite{biau_2015}. With a training set $\mathbb{T}=\left\{\left(\mathbf{x}^{(i)}, \mathbf{y}^{(i)}\right)\right\}$, the KNN algorithm finds the closest $k$ points in the input parameter space. \ignore{This is shown in Figure~\ref{fig:knn visualization} for a new two-dimensional input vector $\mathbf{x}$, which is not part of the training set.} Then, a weighted average is taken over the outputs of the $k$ neighbors to predict the output $\hat{\mathbf{y}}$ of a new input vector $\mathbf{x}$ by 
\begin{equation}
    \hat{\mathbf{y}} = \sum_{i\in\mathrm{NN}(\mathbf{x})} w^{(i)}(\mathbf{x}) \mathbf{y}^{(i)},
\end{equation}
where $\mathrm{NN}(\mathbf{x})$ indicates the $k$ nearest neighbors to $\mathbf{x}.$ Here the weights $w^{(i)}$ is given by the inverse Euclidean distance~\cite{Shepard_ACM_68_ML_interpolation}
\begin{equation}
    w^{(i)}(\mathbf{x}) = \frac{\abs{\mathbf{x} - \mathbf{x}^{(i)}}^{-\alpha}}{\sum_{j\in\mathrm{NN}(\mathbf{x})} \abs{\mathbf{x} - \mathbf{x}^{(j)}}^{-\alpha}}, \quad i \in \mathrm{NN}(\mathbf{x}).
\end{equation}
The optimized hyperparameters learned from training are $k=9, \alpha=5$ for the forward problem and $k=9$, $\alpha=3$ for the inverse problem.

\subsubsection{Feed-Forward Neural Network}

\begin{figure}
\centering
\includegraphics[width=1.0\columnwidth]{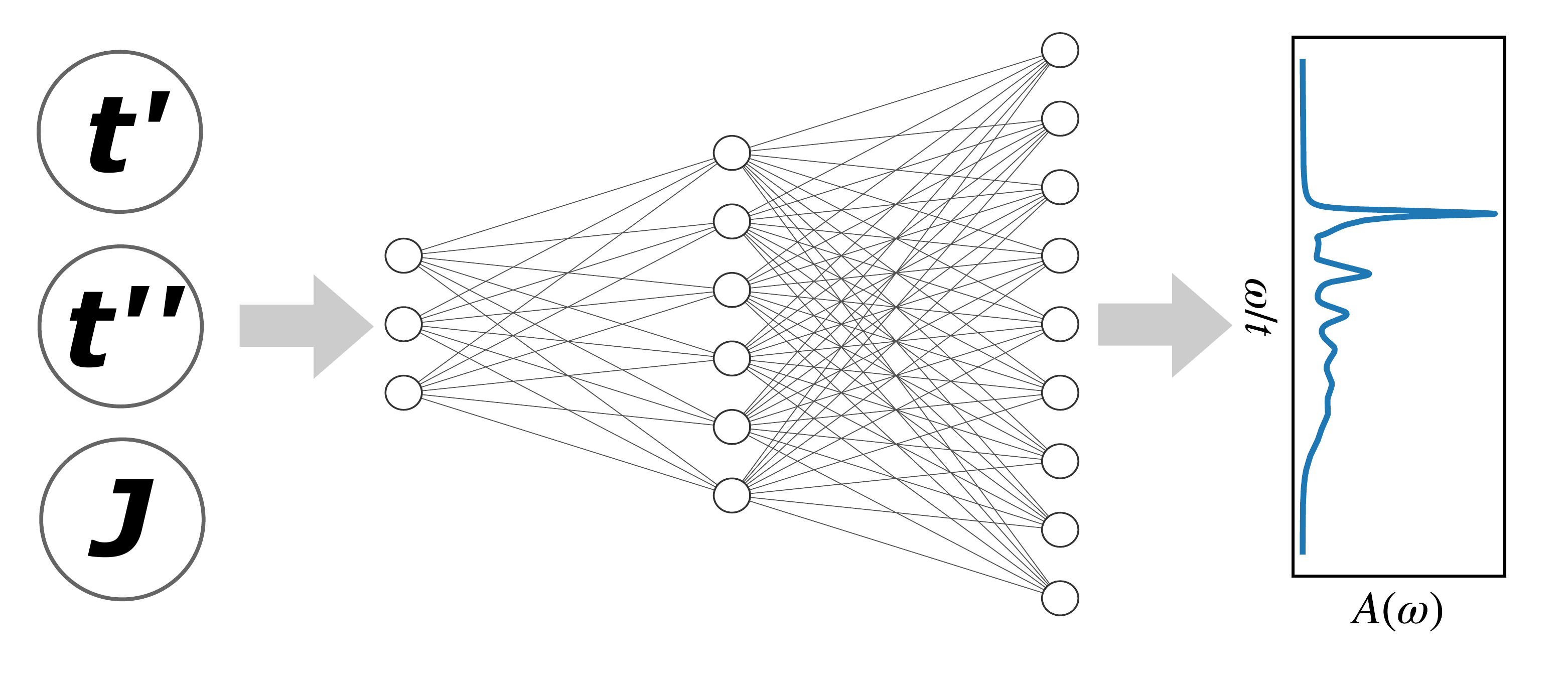}
\caption{\label{fig:fully connected nn} A fully connected neural network, applied to the forward problem of predicting a DOS given ${t', t'', J}$. The neural network takes in a 3 dimensional vector representing ${t', t'', J}$, and outputs a 301 dimensional vector representing the predicted DOS.}
\end{figure}

A neural network is a ML algorithm that implements multiple repeated blocks of linear predictions followed by the application of non-linear functions. In this work, we use feed-forward neural networks (FFNN), which consist of several layers of artificial neurons and is defined by how the layers are implemented and connected. 
Here we use a fully-connected FFNN in which neurons between adjacent layers are fully connected~\cite{Gardner_review_1998_ML_MLP}. For example, a 3-layer FFNN is illustrated in Fig.~\ref{fig:fully connected nn}. The architecture of a fully connected FFNN is primarily defined by the size of each layer, and the layer-by-layer one-way \emph{activation} is given by $\mathbf{a}_l = f_l(W_l \mathbf{a}_{l - 1} + \mathbf{b}_l)$ where $\mathbf{a}_l$ is the $n_l$-dimensional vector output of the $l$th layer, $f_l$ is the activation function, $W_l$ is an $n_l\times n_{l-1}$ matrix of weights ($n_l$ is the number of neurons in layer $l$), and $\mathbf{b}_l$ is a vector of biases. Among them, $W_l$ and $\mathbf{b}_l$ are learned during training. For both the forward and inverse problems, we train the neural networks for 30 minutes, using the rectified linear unit (ReLU) activation function $f_l(x)=\max (0,x)$ and Adam optimizer~\cite{Kingma_arXiv_14_ML_ADAM}.

\section{Results and Discussion}

\subsection{Principal component analysis\label{PCA}}

To analyze the quality of our dataset and visualize the potential of applying ML algorithms to the data, we performed the following principal component analysis (PCA)~\cite{Wold_87_ML_PCA} (see Appendix~\ref{appendix_pca} for details).

\subsubsection{Full DOS data}

\begin{figure}
\centering
\includegraphics[width=1.0\columnwidth]{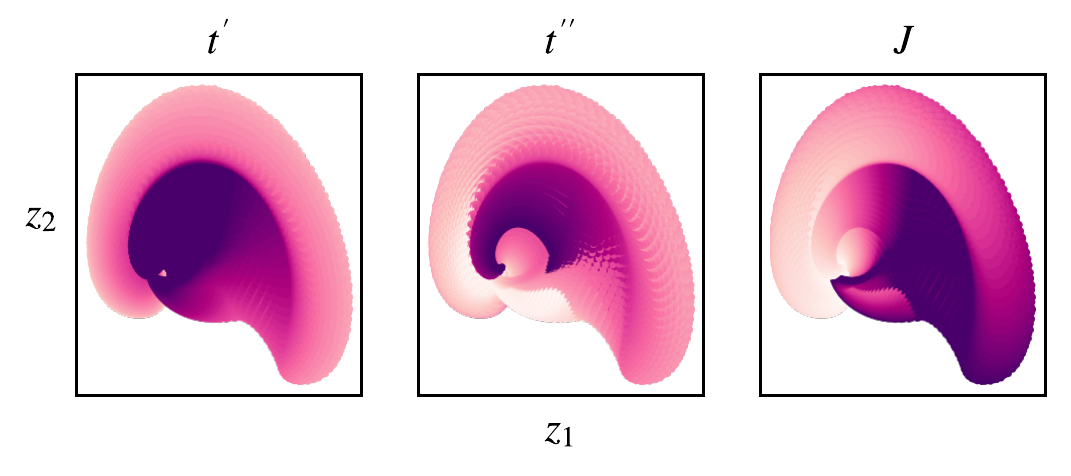}
\caption{\label{fig:pca} 2D visualization of the DOS spectra projected into the first two principal components $(z_1, z_2)^{(i)}$. The color maps of the three subplots are determined by the values of \tp, \tpp, and $J$, respectively. The horizontal and vertical axes represent the first and second principal components, respectively.}
\end{figure}

We proceed with PCA of the dataset in which  $\mathbf{y}^{(i)}=(A(\omega_1),A(\omega_2),\dots,A(\omega_{301}))^{(i)}$ and $\mathbf{x}^{(i)} = (\tpm, \tppm, J)^{(i)}$, where $i=1,2,3,\dots,51^3$.  Following Eq.~(\ref{pca2d}), we show the projected (reduced-dimensional) data vector in Fig.~\ref{fig:pca}, and color each point $(z_1, z_2)^{(i)}$ by the value of $t'^{(i)}$, $t''^{(i)}$, and $J^{(i)}$, respectively, producing three subplots. All results look quite structured (ear like) and the color gradients are smooth. This suggests that the input parameters can be continuously mapped to spectral functions, making ML algorithms well suited for the forward problem. Furthermore, this suggests that the inverse problem of mapping spectral functions to input parameters is feasible. 

\begin{figure}
\centering
\includegraphics[width=1.0\columnwidth]{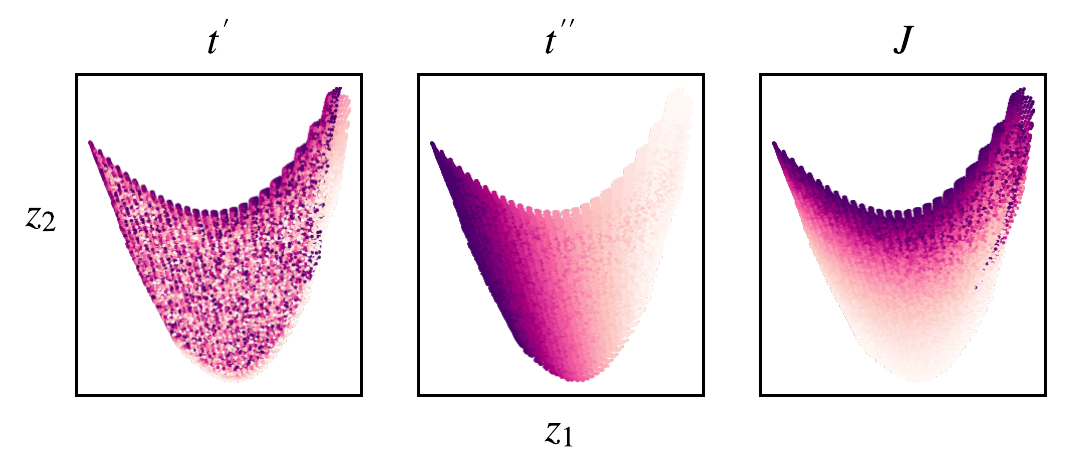}
\caption{\label{fig:lorentzian pca} 2D visualization of the lorentzian parameters (obtained from fitting the first peaks of the DOS spectra) projected into the first two principal components $(z_1, z_2)^{(i)}$. The color maps of the three subplots are determined  by the values of \tp, \tpp, and $J$, respectively.}
\end{figure}

\subsubsection{Using the first peak of DOS}

In comparison, the traditional method for predicting the Hamiltonian parameters is to fit the quasiparticle band $E(\mathbf{k})$ derived from the low-energy peak of the spectral function $A(\mathbf{k},\omega)$, resulting a difficulty determining $t'$~\cite{Yin_PRB_09_WF,Belinicher_PRB_96_Sr2CuO2Cl2}. To visualize this problem with PCA, we use the Lorentzian
\begin{equation}
f(x) = \frac{A \gamma^2}{(x - x_0)^2 + \gamma^2}
\end{equation}
to fit the first peak of every DOS spectrum considered in the inverse problem, resulting in a feature dataset in which now $\mathbf{y}^{(i)} = (A, x_0, \gamma)^{(i)}$. Then, we redo PCA and show the color maps of the first two principal components in Fig.~\ref{fig:lorentzian pca}. We see that while the $t''$ and $J$ plots have smooth gradients, the $t'$ plot contains much more scattering data, demonstrating the difficulty in resolving $t'$ by using only the first-peak information.

\begin{figure}[t]
\centering
\includegraphics[width=1.0\columnwidth]{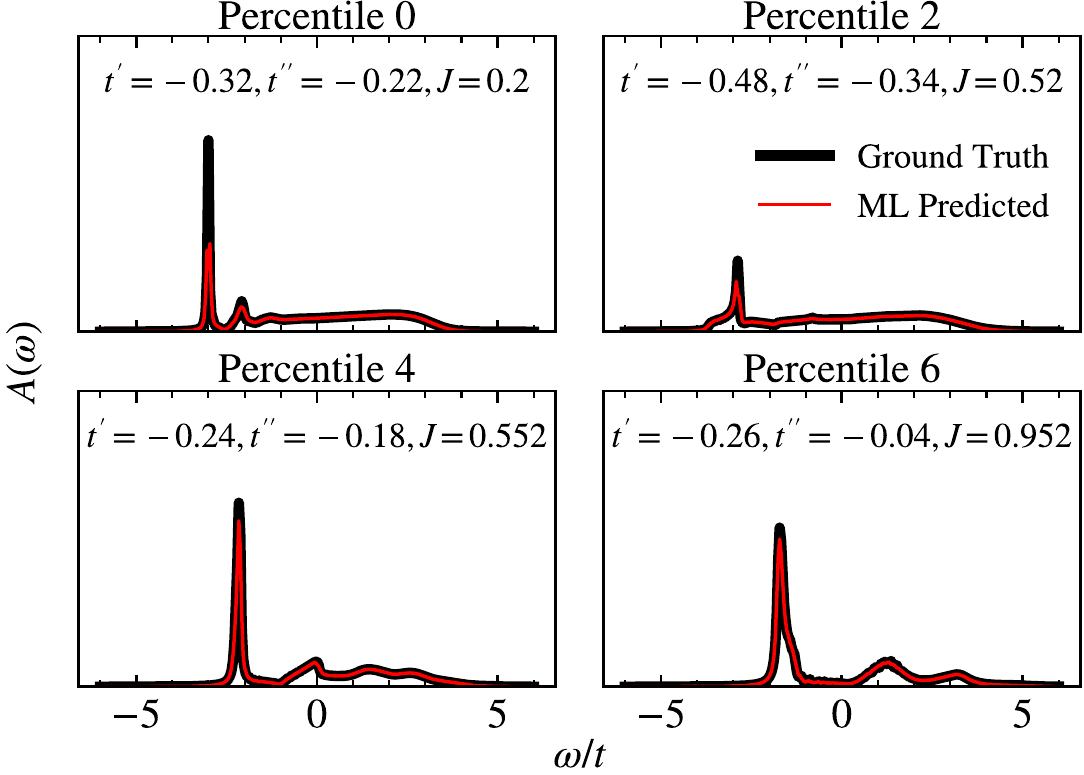}
\vspace{-0.2 in}
\caption{\label{fig:knn percentiles} Comparison of the KNN-predicted DOS and the ground truth (the SCBA-generated DOS) for the worst performing data points in the testing set.}
\end{figure}

\subsection{The forward problem\label{forward}}

\paragraph{KNN.---} We first trained a KNN for the forward problem (see Appendix~\ref{appendix_MSE_forward} for details) and found that with the optimized hyperparameters $k = 9$ and $\alpha = 5$, the KNN was able to  
produce DOS that are almost visually identical to the SCBA results. The worst percentiles of prediction for the testing set, in terms of mean squared error score, are shown in Fig.~\ref{fig:knn percentiles}. We note that even for the examples in the testing set where KNN performs the worst, the KNN prediction is able to reproduce the peak positions and widths almost perfectly, while also performing quite well when reproducing the peak heights.  

\begin{figure}[t]
\centering
\includegraphics[width=1\columnwidth]{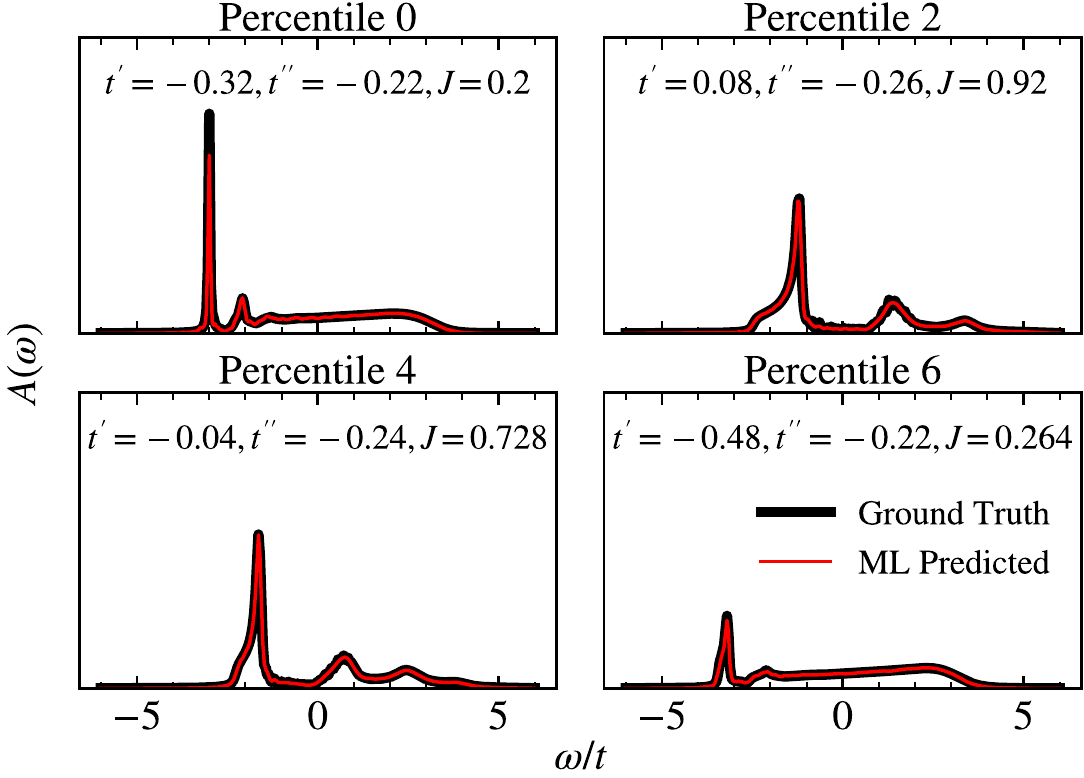}
\vspace{-0.2 in}
\caption{\label{fig:neural net percentiles} Comparison of the neural-network-predicted DOS and the ground truth (the SCBA-generated DOS) for the worst performing data points in the testing set.}
\end{figure}

\paragraph{FFNN.---} We then applied a neural networks for the same task. We found that with optimized hyperparameters $\mathbf{n} = (3, 170, 340, 510, 680, 850, 1020, 301)$, batch size of $1024$, and initial learning rate of $10^{-3}$ (see Appendix~\ref{appendix_MSE_forward} for details), the neural network with six hidden layers was able to 
outperform KNN by roughly a factor of 6 in terms of MSE loss. As shown in Fig.~\ref{fig:neural net percentiles}, even for the worst examples in the testing set, the neural network reproduces peak positions, widths, and heights almost perfectly. For these low-percentile testing examples, the neural network qualitatively appears to reproduce the peak heights better than KNN, which is also manifested quantitatively in its improvement over KNN in terms of MSE loss. 

In addition to excellently reproducing the DOS, the ML algorithms offer a great speedup in computation time over SCBA. While generating the DOS from 130k input combinations using SCBA took over 30 hours, both KNN and the neural network were able to generate the DOS from those same input combinations in seconds: $7.4$ seconds for KNN and $1.2$ seconds for FFNN; KNN and the neural network saw a $1.5 \times 10^4$ and $9 \times 10^4$ speedup over SCBA in predicting the DOS, respectively. 

\begin{table*}
\centering
\caption{\label{tab:Inverse}Examples of worst percentiles when predicting $t'$, $t''$, and $J$ with KNN and FFNN, given the DOS. The numbers in the parentheses are ground truth values.}
\begin{tabular}{c|ccc|ccc}
\hline\hline
 & \multicolumn{3}{c|}{KNN-Predicted} &  \multicolumn{3}{c}{FFNN-Predicted} \\ 
Percentile & $-t'$ & $t''$ & $J$ & $-t'$ & $t''$ & $J$ \\\hline
0 & 0.072(0.02) & 0.130(0.14) & 0.235(0.232) & 0.387(0.38) & 0.497(0.50) & 0.201(0.200)\\
1 & 0.050(0.02) & 0.093(0.10) & 0.565(0.552) & 0.177(0.18) & 0.402(0.40) & 0.889(0.888)\\
2 & 0.235(0.26) & 0.140(0.14) & 0.657(0.664) & 0.022(0.02) & 0.201(0.20) & 0.266(0.264)\\
3 & 0.045(0.02) & 0.440(0.44) & 0.520(0.520) & 0.498(0.50) & 0.481(0.48) & 0.362(0.360)\\
\hline\hline
\end{tabular}
\end{table*}

\subsection{The inverse problem\label{inverse}}

The inverse problem, which involves predicting the model's parameters from observable quantities, has important experimental implications. Since ARPES experiments produce the spectral function and the DOS, the final goal of inverse modeling would be to predict the Hamiltonian parameters from this available experimental data. 
In order to make our DOS dataset more experimentally relevant, we shifted every DOS with respect to the top of the quasiparticle valence band [see Fig.~\ref{fig:problem}(b)]. This better mimics experimental data, where absolute energies are not measured, but are instead found relative to the Fermi level [see Fig.~\ref{fig:problem}(a)]. We also limited the dataset to include only examples with $t' < 0$ and $t'' > 0$, i.e, the hole doped case (the case of $t' > 0$ and $t'' > 0$ corresponds to electron doping). As different DOS in our dataset requires shifting of different amounts, this demands an expansion of our energy window. As a result, we now use a 354-point linear grid to sample the shifted DOS. The input is $\mathbf{x}^{(i)}=(A(\omega_1),A(\omega_2),\dots,A(\omega_{354}))^{(i)}$ and the output is $\mathbf{y}^{(i)} = (\tpm, \tppm, J)^{(i)}$, where $i=1,2,3,\dots,\sim 51^3/4$. 

\paragraph{KNN.---} We first use a KNN to predict the corresponding $t'$, $t''$, and $J$, given a DOS. With the trained hyperparameters $k = 9$ and $\alpha = 3$, the results for the worst-percentile predictions are displayed in Table~\ref{tab:Inverse}. 
We see that for worst percentiles, a KNN is able to predict $t''$ and $J$ quite accurately, but has trouble with $t'$. This means that the outstanding problem of predicting $t'$ likely cannot be resolved by this classic modeling method.

\paragraph{FFNN.---} We then trained a FFNN for the same task. We found that with the hyperparameters $\mathbf{n} = (301, 256, 128, 64, 32, 3)$, batch size of $128$, and a learning rate of $10^{-3}$  (see Appendix~\ref{appendix_MSE_inverse} for details), the neural network is able to significantly outperform KNN, with the MSE being $67\times$ better than that of KNN. 
As shown in Table~\ref{tab:Inverse}, 
even for the worst examples in the test set, the neural network can predict at least the first two significant figures. Thus, the neural network offers an accurate approach to prediction of material parameters. 

\subsection{Inverse problem: Finding $t$\label{inverse_t}}

\begin{figure}[b]
\centering
\includegraphics[width=1.0\columnwidth]{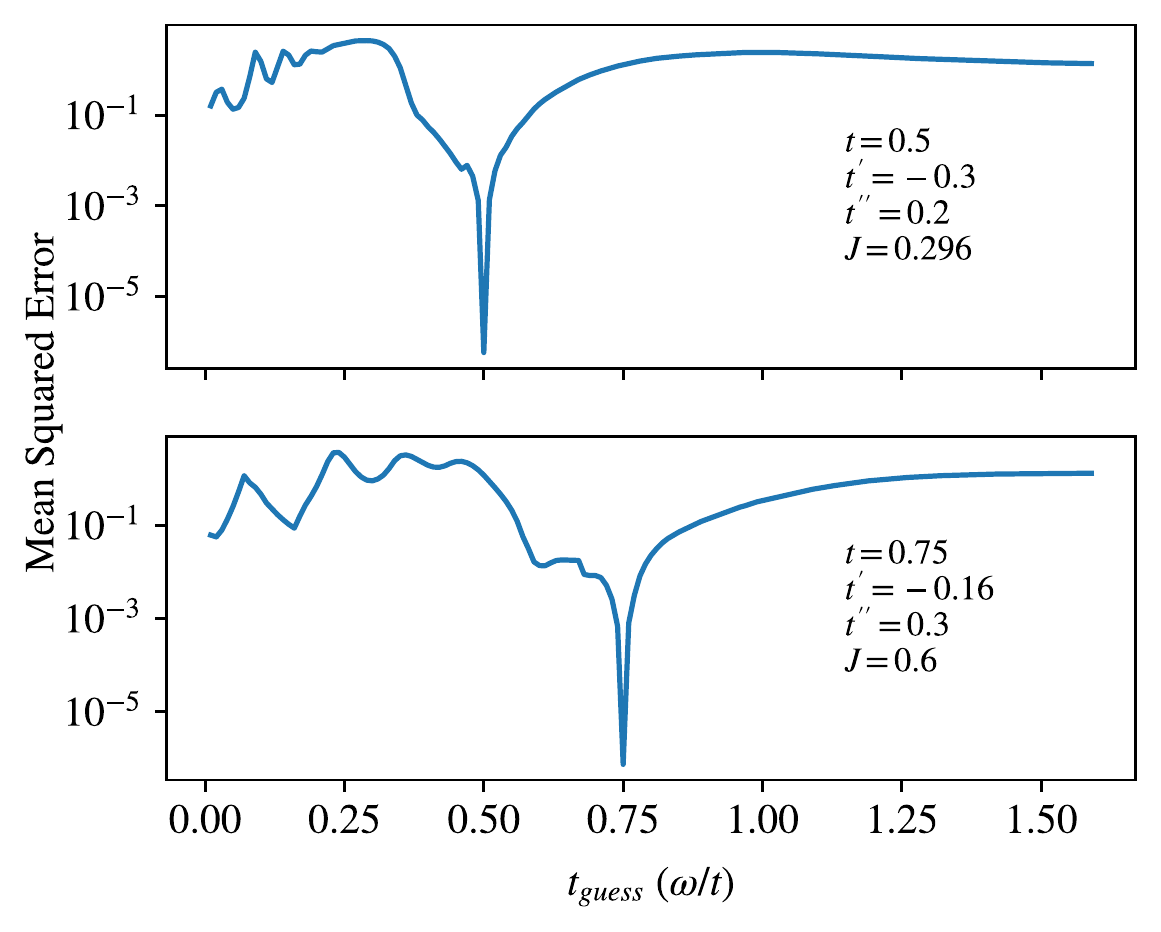}
\caption{\label{fig:t prediction} The mean squared error when rescaling a mock ``experimental'' DOS in units of $t_\mathrm{guess}$, and running FFNN to predict the ground truth $t'$, $t''$, and $J$.}. 
\end{figure}

In our above analysis, we have produced and analyzed DOS with $t$ being the energy unit, i.e., $t = 1$. However, ARPES experiments produce DOS that are measured in terms of absolute energy. We thus proceed to analyze the feasibility of obtaining ground truth $t$ (referred to as $t_\mathrm{truth}$) from more experimentally realistic DOS. To this end, we examine the following simulation: We start with an SCBA-generated DOS with $t = 1$ from our dataset, shifted with respect to the top of the valence band. We then scale the DOS by using $A(\omega) \to A(\omega\,t_\mathrm{truth})/t_\mathrm{truth}$, thus producing an ``experimental'' DOS in units of absolute energy. The task is to find $t_\mathrm{truth}$ from this ``experimental'' DOS while pretending that we do not know this $t_\mathrm{truth}$. 

We propose the following algorithm for this task: We add to the ML methods presented in Section~\ref{inverse} an outermost loop over various guesses of $t_\mathrm{truth}$. Specifically, for each $t_\mathrm{guess}$, we rescale the experimental DOS by using $A(\omega) \to A(\omega/t_\mathrm{guess})\,t_\mathrm{guess}$, producing the DOS with $t_\mathrm{guess}$ being the energy unit; thus, we arrive at the same inverse problem of predicting $t'$, $t''$, and $J$ with $t=1$ studied in Section~\ref{inverse}. Then, we resample the rescale DOS with the the same 354 point energy grid as for the previous inverse problem, using cubic spline interpolation. After that, we run the trained neural network to produce $t'$, $t''$, and $J$ from the rescaled DOS and calculate the MSE.

The results of this procedure are shown in Fig.~\ref{fig:t prediction}. We find that this algorithm is able to predict $t_{truth}$ very accurately, as seen by the steep drop in the mean squared error when $t_\mathrm{guess} = t_\mathrm{truth}$. Adding such one outermost loop takes advantage of the scalability of the spectral functions [Eq.~(\ref{eq:scale})] and reduces the dimensionality of the model parameter space from 4 to 3, a significant improvement in coping with the curse of dimensionality in big-data ML research.

\section{Summary\label{Summary}}
We have investigated the potential of ML algorithms for understanding the spectral functions of a hole in the {\tttJ} model and found that ML algorithms are well suited for the task. The analysis of the dataset of SCBA-generated spectral functions demonstrates the presence of a continuous mapping between the model parameters and the resulting DOS. Given a set of the model parameters, we found that both KNN and neural networks can produce almost visually identical DOS as SCBA, with a speedup of as much as $9 \times 10^4$. We also found that the ML algorithms, especially deep learning neural networks, can predict $t'$, $t''$, and $J$ very accurately given a DOS. With such a speedup in the calculation of DOS, as well as the ability to solve the inverse problem, ML offers a potential tool to search for the model parameters that produce desirable spectral functions. The present method can be directly applied to other cases of energy distribution curves (EDC) such as $A(\mathbf{k},\omega)$ at constant momentum or the cases of momentum distribution curves (MDC), which are the intensities as a function of momentum at constant energy~\cite{Valla_Science_99_MDC}. Future work will focus on working with  experimental data, which are further complicated by instrument resolution and irreducible noise.

\section*{Data Availability}
The data generated and used in this study are openly available from the Zenodo database~\cite{lee_jackson_2023_7527378}.

\ignore{
\section*{Code Availability}
The Pytorch Lightning research framework used in this work is publicly available. **We shall decide later on whether our ML code will be made available to the public.** 
}

\begin{acknowledgements}
This work was supported by U.S. Department of Energy (DOE) the Office of Science, Office of Basic Energy Sciences, Materials Sciences and Engineering Division under Contract No. DE-SC0012704. This project was supported in part by the U.S. Department of Energy, Office of Science, Office of Workforce Development for Teachers and Scientists (WDTS) under the Science Undergraduate Laboratory Internships Program (SULI). This project was supported in part by the Brookhaven National Laboratory (BNL), Condensed Matter Physics and Materials Science Division under the BNL Supplemental Undergraduate Research Program (SURP).
\end{acknowledgements}

\ignore{
\section*{Author Contributions}
W.Y. designed the project and wrote the SCBA code. J.L. wrote the ML code with the guidance of M.R.C. and performed the calculations. All authors contributed to the analysis and discussion of the results. J.L. and W.Y. wrote the manuscript, with M.R.C. contributing significant edits and revisions.

\section*{COMPETING INTERESTS}
The authors declare no competing interests.

\section*{ADDITIONAL INFORMATION}
\textbf{Correspondence} and requests for materials should be addressed to Weiguo Yin.

}

\appendix

\section{The self-consistent Born approximation\label{appendix_scba}}

SCBA uses non-crossing Feynman diagrams to calculate Green’s function $G(\mathbf{k}, \omega)$ used in Eqs.~(\ref{eq:A}) and (\ref{eq:G}), which describes the propagation of a particle in the lattice. The self-consistent system of equations to be solved is 
\begin{eqnarray}
G(\mathbf{k}, \omega)&=&\left[G^0(\mathbf{k}, \omega)^{-1}-\Sigma(\mathbf{k}, \omega)\right]^{-1}, \\
    \Sigma(\mathbf{k}, \omega)&=&\sum_{\mathbf{q}}{|M(\mathbf{k},\mathbf{q})|^2 G(\mathbf{k-q}, \omega)},
\end{eqnarray}
where $\Sigma(\mathbf{k}, \omega)$ is the so-called self-energy, $G^0(\mathbf{k}, \omega)= \lim_{\eta \to 0^+}{\left[\omega+i\eta-\epsilon_\mathbf{k}\right]^{-1}}$ is the bare Green’s function with $\epsilon_\mathbf{k}=4\tpm \cos k_x \cos k_y + 2\tppm [\cos (2k_x) + \cos (2k_y)]$ being the bare dispersion relation of the hole quasiparticle.


In order to generate a high-quality dataset of spectral functions, SCBA samples over a dense mesh for both the hole momentum $\mathbf{k}$ and the magnon momentum $\mathbf{q}$. The sizes of $\mathbf{k}$ and $\mathbf{q}$ can be different while being commensurate, corresponding to the application of twisted boundary conditions~\cite{Yin_PRB_09_FTBC}. While higher density k and q sampling leads to higher quality spectral functions, they are also more computationally expensive. We tested various combinations of $\mathbf{k}$ sampling density and $\mathbf{q}$ sampling density. We found that above sampling densities of a $128 \times 128$ lattice for $\kbf$, and a $32 \times 32$ lattice for $\qbf$, the results converge. 


\section{Principal component analysis\label{appendix_pca}}

Given the training set $\mathbb{T}=\left\{\left(\mathbf{x}^{(i)}, \mathbf{y}^{(i)}\right)\right\}$ of size $N$ where $\mathbf{x}^{(i)}$ is an $n$-dimensional vector and $\mathbf{y}^{(i)}$ is a $m$-dimensional vector, we begin with the $m \times N$ matrix $Z =\left(\textbf{y}^{(1)},\textbf{y}^{(2)},\dots,\textbf{y}^{(N)}\right)$ with each column representing a $\mathbf{y}^{(i)}$. The $m$ rows of $Z$ are then each shifted so that the mean of every raw is zero, that is, the center of the data is translated to the origin of the $m$-dimensional space, which does not change how the data points are positioned relative to each other. The $m\times m$ covariance matrix is given by
\begin{equation}
\mathbf{C_Z} = \frac{1}{m}\mathbf{Z}\mathbf{Z}^\mathrm{T}
\end{equation}
The principal components of $\mathbf{C_Z}$ are just its normalized eigenvectors $\mathbf{e}_j$ with $j=1,2,3,\dots,m$, arranged according to their corresponding eigenvalues (variance) in descending order. $\mathbf{e}_1$ is the direction in the $m$-dimensional space with the largest variance in $Z$,  $\mathbf{e}_2$ is the direction with the second largest variance, and so on. One can use the eigenvalues to determine the proportion of the variation that each principal component accounts for. If $\mathbf{e}_1$ and $\mathbf{e}_2$ account for the vast majority of the variation in the data, a 2D graph, using only $\mathbf{e}_1$ and $\mathbf{e}_2$ as the axes, would be a good approximation of an unimaginable $m$-dimensional graph. The coordinates of $\mathbf{y}^{(i)}$ projected into the 2D subspace is given by \begin{equation}
\label{pca2d}
    \mathbf{z}^{(i)}\equiv (z_1,z_2)^{(i)}=(\mathbf{e}_1 \cdot \mathbf{y}^{(i)}, \mathbf{e}_2\cdot \mathbf{y}^{(i)}).
\end{equation}
Then, a 2D color map can be produced by coloring all the $\mathbf{z}^{(i)}$ points according to the value of an element in the vector $\mathbf{x}^{(i)}$, so we can obtain $n$ such 2D color maps. Among them, those appearing to be quite structured and smooth in the color gradients suggesting that the corresponding input parameters can be continuously mapped to spectral functions, making ML algorithms well suited for the task. This also suggests that the inverse problem of mapping spectral functions to input parameters is also feasible. 

\section{Total mean squared error}
\subsection{MSE for the forward problem\label{appendix_MSE_forward}}

\begin{figure}[t]
\centering
\includegraphics[width=1.0\columnwidth]{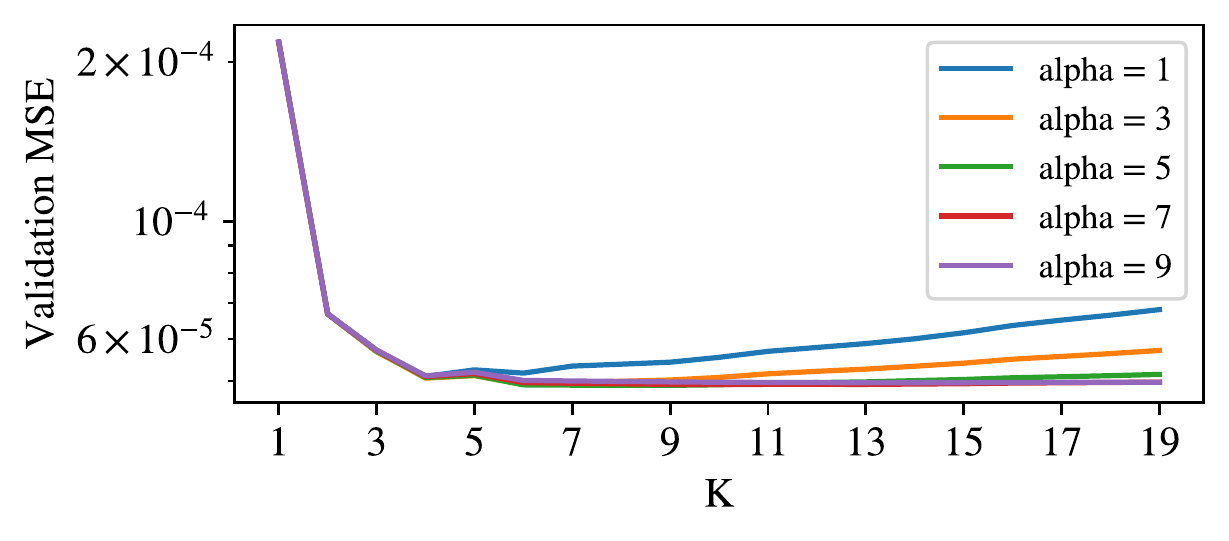}
\vspace{-0.3in}
\caption{\label{fig:knn hyperparameter tuning} Validation loss for the forward problem for different values of k, and $\alpha$.}
\end{figure}

Hyperparameter tuning was performed by optimizing hyperparameters to reduce validation set error. 

For KNN, we tested various values of $\alpha$ and $k$ via grid search. Fig.~\ref{fig:knn hyperparameter tuning} shows a plot of the validation mean squared error for various values of $k$, and $\alpha$, including the optimal value $\alpha = 5$ and $k = 9$.

\begin{figure}[t]
\centering
\includegraphics[width=\columnwidth]{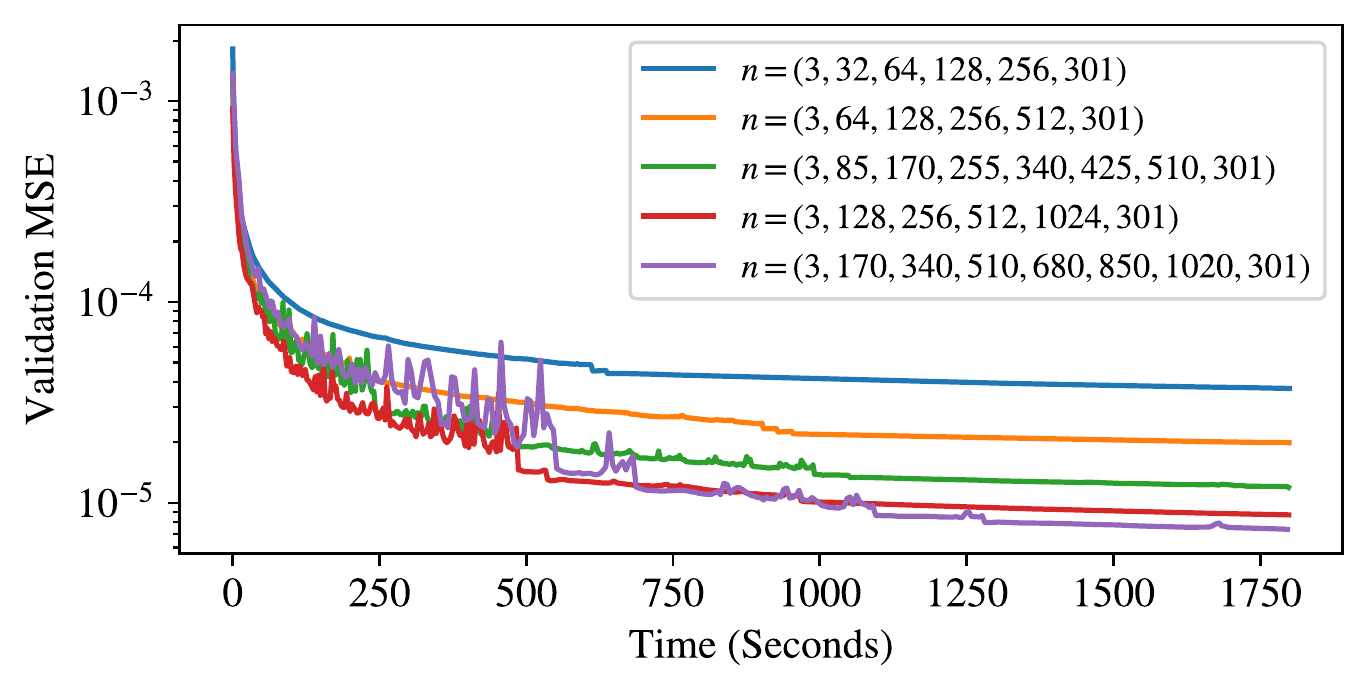}
\vspace{-0.2in}
\caption{\label{fig:FFNN Validation vs Time} Validation loss over time for various different FFNN architectures, with \texttt{bs} $ = 1024$, and \texttt{lr} $ = 10^{-3}$, for the forward problem. 
}
\end{figure}

For FFNN, we tuned hyperparameters with a combination of hand tuning and grid search. The architecture of the neural network $\mathbf{n}$ was of particular interest in hyperparameter tuning. We tested a variety of architectures with different number of layers, which ``ramped'' up to a different number of neurons in the final hidden layer. Fig.~\ref{fig:FFNN Validation vs Time} shows a plot of the the validation error over time for different architectures over a 30 minute time period. 

\begin{figure}[h]
\centering
\includegraphics[width=1\columnwidth]{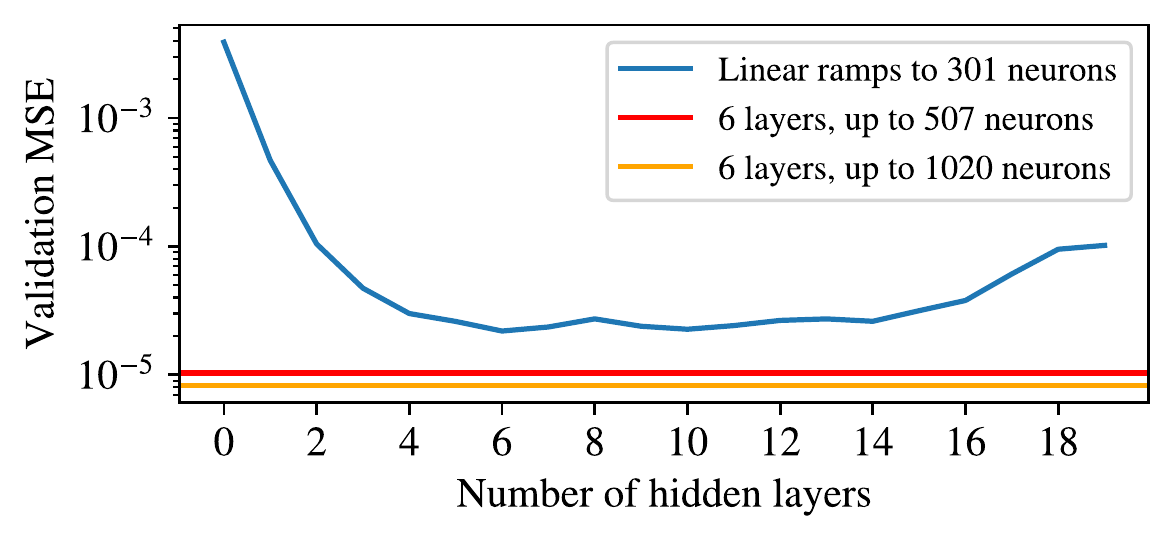}
\vspace{-0.2in}
\caption{\label{fig:FFNN Linear Ramps} Validation loss for the forward problem for linear ramps with different number of hidden layers. These are compared to architectures that include more than 301 neurons in the hidden layers.}
\end{figure}

For the architecture design, we tested various ``linear ramps'' which simply ramp from 3 input neurons to the 301 output neurons linearly. For example, a linear-ramp architecture with 3 hidden layers would have $\mathbf{n} = (3, 77, 151, 225, 301)$. We found that while these linear ramps intuitively made more sense, they trained considerably less efficiently than architectures which had more than 301 neurons in the hidden layers. We see in Fig.~\ref{fig:FFNN Linear Ramps} that after training for 30 minutes, even the best linear ramps perform worse on the validation set than architectures which include larger hidden layers. 

Other hyperparameters include the batch size and the learning rate. The batch size (\texttt{bs}) is the size of a subset of $\mathbb{T}$ fed to the neural network to perform a single gradient update. One epoch of training completes after all the training data have been fed through the network (in a randomized order each time). The learning rate (\texttt{lr}) is the base step size for tuning weights towards the optimization direction (along gradient descent) and is scheduled to decrease by a factor of 2 when no improvement is realized after 10 epochs. For the forward problem, we find optimized hyperparameters $\mathbf{n} = (3, 170, 340, 510, 680, 850, 1020, 301)$, \texttt{bs} $ = 1024$, and \texttt{lr} $ = 10^{-3}$. 

After optimizing hyperparameters using the validation set for both KNN and FFNN, the following MSE results are derived from the performance of the ML models on the testing set: $4.71\times10^{-5}$ for KNN and $7.24\times10^{-6}$ for FFNN.

\vspace{0.2 in}

\subsection{MSE for the inverse problem\label{appendix_MSE_inverse}}
For KNN, we again used grid search to tune hyperparameters.

For FFNN, while we found that an architecture of $\mathbf{n} = (354, 256, 128, 64, 32, 3)$ together with \texttt{bs} $ = 128$ and \texttt{lr} $ = 10^{-3}$ after hyperparameter tuning, we note that several architectures, which ramped down from 354 to 3 neurons, performed similarly on the validation set.

After optimizing hyperparameters using the validation set for both KNN and FFNN, the following MSE results are derived from the performance of the ML models on the testing set: $4.19\times10^{-5}$ for KNN and $6.29\times10^{-7}$ for FFNN.


\input{ML_t-J.bbl}

\end{document}

%% file: ML_t-J.bbl
%

%% file: main.bbl
\begin{thebibliography}{72}%
\makeatletter
\providecommand \@ifxundefined [1]{%
 \@ifx{#1\undefined}
}%
\providecommand \@ifnum [1]{%
 \ifnum #1\expandafter \@firstoftwo
 \else \expandafter \@secondoftwo
 \fi
}%
\providecommand \@ifx [1]{%
 \ifx #1\expandafter \@firstoftwo
 \else \expandafter \@secondoftwo
 \fi
}%
\providecommand \natexlab [1]{#1}%
\providecommand \enquote  [1]{``#1''}%
\providecommand \bibnamefont  [1]{#1}%
\providecommand \bibfnamefont [1]{#1}%
\providecommand \citenamefont [1]{#1}%
\providecommand \href@noop [0]{\@secondoftwo}%
\providecommand \href [0]{\begingroup \@sanitize@url \@href}%
\providecommand \@href[1]{\@@startlink{#1}\@@href}%
\providecommand \@@href[1]{\endgroup#1\@@endlink}%
\providecommand \@sanitize@url [0]{\catcode `\\12\catcode `\$12\catcode
  `\&12\catcode `\#12\catcode `\^12\catcode `\_12\catcode `\%12\relax}%
\providecommand \@@startlink[1]{}%
\providecommand \@@endlink[0]{}%
\providecommand \url  [0]{\begingroup\@sanitize@url \@url }%
\providecommand \@url [1]{\endgroup\@href {#1}{\urlprefix }}%
\providecommand \urlprefix  [0]{URL }%
\providecommand \Eprint [0]{\href }%
\providecommand \doibase [0]{http://dx.doi.org/}%
\providecommand \selectlanguage [0]{\@gobble}%
\providecommand \bibinfo  [0]{\@secondoftwo}%
\providecommand \bibfield  [0]{\@secondoftwo}%
\providecommand \translation [1]{[#1]}%
\providecommand \BibitemOpen [0]{}%
\providecommand \bibitemStop [0]{}%
\providecommand \bibitemNoStop [0]{.\EOS\space}%
\providecommand \EOS [0]{\spacefactor3000\relax}%
\providecommand \BibitemShut  [1]{\csname bibitem#1\endcsname}%
\let\auto@bib@innerbib\@empty
\bibitem [{\citenamefont {Dagotto}(1994)}]{Dagotto_RMP_94}%
  \BibitemOpen
  \bibfield  {author} {\bibinfo {author} {\bibfnamefont {E.}~\bibnamefont
  {Dagotto}},\ }\href {\doibase 10.1103/RevModPhys.66.763} {\bibfield
  {journal} {\bibinfo  {journal} {Rev. Mod. Phys.}\ }\textbf {\bibinfo {volume}
  {66}},\ \bibinfo {pages} {763} (\bibinfo {year} {1994})}\BibitemShut
  {NoStop}%
\bibitem [{\citenamefont {Lee}\ \emph {et~al.}(2006)\citenamefont {Lee},
  \citenamefont {Nagaosa},\ and\ \citenamefont {Wen}}]{Lee_RMP_06}%
  \BibitemOpen
  \bibfield  {author} {\bibinfo {author} {\bibfnamefont {P.~A.}\ \bibnamefont
  {Lee}}, \bibinfo {author} {\bibfnamefont {N.}~\bibnamefont {Nagaosa}}, \ and\
  \bibinfo {author} {\bibfnamefont {X.-G.}\ \bibnamefont {Wen}},\ }\href
  {\doibase 10.1103/RevModPhys.78.17} {\bibfield  {journal} {\bibinfo
  {journal} {Rev. Mod. Phys.}\ }\textbf {\bibinfo {volume} {78}},\ \bibinfo
  {pages} {17} (\bibinfo {year} {2006})}\BibitemShut {NoStop}%
\bibitem [{\citenamefont {Schmitt-Rink}\ \emph {et~al.}(1988)\citenamefont
  {Schmitt-Rink}, \citenamefont {Varma},\ and\ \citenamefont
  {Ruckenstein}}]{Schmitt-Rink_PRL_88_SCBA}%
  \BibitemOpen
  \bibfield  {author} {\bibinfo {author} {\bibfnamefont {S.}~\bibnamefont
  {Schmitt-Rink}}, \bibinfo {author} {\bibfnamefont {C.~M.}\ \bibnamefont
  {Varma}}, \ and\ \bibinfo {author} {\bibfnamefont {A.~E.}\ \bibnamefont
  {Ruckenstein}},\ }\href {\doibase 10.1103/PhysRevLett.60.2793} {\bibfield
  {journal} {\bibinfo  {journal} {Phys. Rev. Lett.}\ }\textbf {\bibinfo
  {volume} {60}},\ \bibinfo {pages} {2793} (\bibinfo {year}
  {1988})}\BibitemShut {NoStop}%
\bibitem [{\citenamefont {Johnson}\ \emph {et~al.}(2015)\citenamefont
  {Johnson}, \citenamefont {Xu},\ and\ \citenamefont {Yin}}]{Yin_book_15}%
  \BibitemOpen
  \bibinfo {editor} {\bibfnamefont {P.~D.}\ \bibnamefont {Johnson}}, \bibinfo
  {editor} {\bibfnamefont {G.}~\bibnamefont {Xu}}, \ and\ \bibinfo {editor}
  {\bibfnamefont {W.-G.}\ \bibnamefont {Yin}},\ eds.,\ \href {\doibase
  10.1007/978-3-319-11254-1} {\emph {\bibinfo {title} {Iron-Based
  Superconductivity}}},\ \bibinfo {series} {Springer Series in Materials
  Science}, Vol.\ \bibinfo {volume} {211}\ (\bibinfo  {publisher} {Springer
  International Publishing Switzerland},\ \bibinfo {year} {2015})\BibitemShut
  {NoStop}%
\bibitem [{\citenamefont {Yin}\ \emph {et~al.}(2010)\citenamefont {Yin},
  \citenamefont {Lee},\ and\ \citenamefont {Ku}}]{Yin_PRL_10_FeTe}%
  \BibitemOpen
  \bibfield  {author} {\bibinfo {author} {\bibfnamefont {W.-G.}\ \bibnamefont
  {Yin}}, \bibinfo {author} {\bibfnamefont {C.-C.}\ \bibnamefont {Lee}}, \ and\
  \bibinfo {author} {\bibfnamefont {W.}~\bibnamefont {Ku}},\ }\href {\doibase
  10.1103/PhysRevLett.105.107004} {\bibfield  {journal} {\bibinfo  {journal}
  {Phys. Rev. Lett.}\ }\textbf {\bibinfo {volume} {105}},\ \bibinfo {pages}
  {107004} (\bibinfo {year} {2010})}\BibitemShut {NoStop}%
\bibitem [{\citenamefont {Cao}\ \emph {et~al.}(2018)\citenamefont {Cao},
  \citenamefont {Fatemi}, \citenamefont {Fang}, \citenamefont {Watanabe},
  \citenamefont {Taniguchi}, \citenamefont {Kaxiras},\ and\ \citenamefont
  {Jarillo-Herrero}}]{Cao_Nature_18_TBG_SC}%
  \BibitemOpen
  \bibfield  {author} {\bibinfo {author} {\bibfnamefont {Y.}~\bibnamefont
  {Cao}}, \bibinfo {author} {\bibfnamefont {V.}~\bibnamefont {Fatemi}},
  \bibinfo {author} {\bibfnamefont {S.}~\bibnamefont {Fang}}, \bibinfo {author}
  {\bibfnamefont {K.}~\bibnamefont {Watanabe}}, \bibinfo {author}
  {\bibfnamefont {T.}~\bibnamefont {Taniguchi}}, \bibinfo {author}
  {\bibfnamefont {E.}~\bibnamefont {Kaxiras}}, \ and\ \bibinfo {author}
  {\bibfnamefont {P.}~\bibnamefont {Jarillo-Herrero}},\ }\href
  {https://doi.org/10.1038/nature26160} {\bibfield  {journal} {\bibinfo
  {journal} {Nature}\ }\textbf {\bibinfo {volume} {556}},\ \bibinfo {pages}
  {43} (\bibinfo {year} {2018})}\BibitemShut {NoStop}%
\bibitem [{\citenamefont {Ji}\ \emph {et~al.}(2021)\citenamefont {Ji},
  \citenamefont {Xu}, \citenamefont {Kendrick}, \citenamefont {Chiu},
  \citenamefont {Br\"uggenj\"urgen}, \citenamefont {Greif}, \citenamefont
  {Bohrdt}, \citenamefont {Grusdt}, \citenamefont {Demler}, \citenamefont
  {Lebrat},\ and\ \citenamefont {Greiner}}]{Ji_PRX_21}%
  \BibitemOpen
  \bibfield  {author} {\bibinfo {author} {\bibfnamefont {G.}~\bibnamefont
  {Ji}}, \bibinfo {author} {\bibfnamefont {M.}~\bibnamefont {Xu}}, \bibinfo
  {author} {\bibfnamefont {L.~H.}\ \bibnamefont {Kendrick}}, \bibinfo {author}
  {\bibfnamefont {C.~S.}\ \bibnamefont {Chiu}}, \bibinfo {author}
  {\bibfnamefont {J.~C.}\ \bibnamefont {Br\"uggenj\"urgen}}, \bibinfo {author}
  {\bibfnamefont {D.}~\bibnamefont {Greif}}, \bibinfo {author} {\bibfnamefont
  {A.}~\bibnamefont {Bohrdt}}, \bibinfo {author} {\bibfnamefont
  {F.}~\bibnamefont {Grusdt}}, \bibinfo {author} {\bibfnamefont
  {E.}~\bibnamefont {Demler}}, \bibinfo {author} {\bibfnamefont
  {M.}~\bibnamefont {Lebrat}}, \ and\ \bibinfo {author} {\bibfnamefont
  {M.}~\bibnamefont {Greiner}},\ }\href {\doibase 10.1103/PhysRevX.11.021022}
  {\bibfield  {journal} {\bibinfo  {journal} {Phys. Rev. X}\ }\textbf {\bibinfo
  {volume} {11}},\ \bibinfo {pages} {021022} (\bibinfo {year}
  {2021})}\BibitemShut {NoStop}%
\bibitem [{\citenamefont {Koepsell}\ \emph {et~al.}(2021)\citenamefont
  {Koepsell}, \citenamefont {Bourgund}, \citenamefont {Sompet}, \citenamefont
  {Hirthe}, \citenamefont {Bohrdt}, \citenamefont {Wang}, \citenamefont
  {Grusdt}, \citenamefont {Demler}, \citenamefont {Salomon}, \citenamefont
  {Gross},\ and\ \citenamefont {Bloch}}]{Koepsell_Science_21_cold-atom}%
  \BibitemOpen
  \bibfield  {author} {\bibinfo {author} {\bibfnamefont {J.}~\bibnamefont
  {Koepsell}}, \bibinfo {author} {\bibfnamefont {D.}~\bibnamefont {Bourgund}},
  \bibinfo {author} {\bibfnamefont {P.}~\bibnamefont {Sompet}}, \bibinfo
  {author} {\bibfnamefont {S.}~\bibnamefont {Hirthe}}, \bibinfo {author}
  {\bibfnamefont {A.}~\bibnamefont {Bohrdt}}, \bibinfo {author} {\bibfnamefont
  {Y.}~\bibnamefont {Wang}}, \bibinfo {author} {\bibfnamefont {F.}~\bibnamefont
  {Grusdt}}, \bibinfo {author} {\bibfnamefont {E.}~\bibnamefont {Demler}},
  \bibinfo {author} {\bibfnamefont {G.}~\bibnamefont {Salomon}}, \bibinfo
  {author} {\bibfnamefont {C.}~\bibnamefont {Gross}}, \ and\ \bibinfo {author}
  {\bibfnamefont {I.}~\bibnamefont {Bloch}},\ }\href {\doibase
  10.1126/science.abe7165} {\bibfield  {journal} {\bibinfo  {journal}
  {Science}\ }\textbf {\bibinfo {volume} {374}},\ \bibinfo {pages} {82}
  (\bibinfo {year} {2021})}\BibitemShut {NoStop}%
\bibitem [{\citenamefont {Koepsell}\ \emph {et~al.}(2019)\citenamefont
  {Koepsell}, \citenamefont {Vijayan}, \citenamefont {Sompet}, \citenamefont
  {Grusdt}, \citenamefont {Hilker}, \citenamefont {Demler}, \citenamefont
  {Salomon}, \citenamefont {Bloch},\ and\ \citenamefont
  {Gross}}]{Koepsell_Nature_19_cold-atom}%
  \BibitemOpen
  \bibfield  {author} {\bibinfo {author} {\bibfnamefont {J.}~\bibnamefont
  {Koepsell}}, \bibinfo {author} {\bibfnamefont {J.}~\bibnamefont {Vijayan}},
  \bibinfo {author} {\bibfnamefont {P.}~\bibnamefont {Sompet}}, \bibinfo
  {author} {\bibfnamefont {F.}~\bibnamefont {Grusdt}}, \bibinfo {author}
  {\bibfnamefont {T.~A.}\ \bibnamefont {Hilker}}, \bibinfo {author}
  {\bibfnamefont {E.}~\bibnamefont {Demler}}, \bibinfo {author} {\bibfnamefont
  {G.}~\bibnamefont {Salomon}}, \bibinfo {author} {\bibfnamefont
  {I.}~\bibnamefont {Bloch}}, \ and\ \bibinfo {author} {\bibfnamefont
  {C.}~\bibnamefont {Gross}},\ }\href
  {https://doi.org/10.1038/s41586-019-1463-1} {\bibfield  {journal} {\bibinfo
  {journal} {Nature}\ }\textbf {\bibinfo {volume} {572}},\ \bibinfo {pages}
  {358} (\bibinfo {year} {2019})}\BibitemShut {NoStop}%
\bibitem [{\citenamefont {Bohrdt}\ \emph {et~al.}(2019)\citenamefont {Bohrdt},
  \citenamefont {Chiu}, \citenamefont {Ji}, \citenamefont {Xu}, \citenamefont
  {Greif}, \citenamefont {Greiner}, \citenamefont {Demler}, \citenamefont
  {Grusdt},\ and\ \citenamefont {Knap}}]{Bohrdt_NP_19_ML_snapshots}%
  \BibitemOpen
  \bibfield  {author} {\bibinfo {author} {\bibfnamefont {A.}~\bibnamefont
  {Bohrdt}}, \bibinfo {author} {\bibfnamefont {C.~S.}\ \bibnamefont {Chiu}},
  \bibinfo {author} {\bibfnamefont {G.}~\bibnamefont {Ji}}, \bibinfo {author}
  {\bibfnamefont {M.}~\bibnamefont {Xu}}, \bibinfo {author} {\bibfnamefont
  {D.}~\bibnamefont {Greif}}, \bibinfo {author} {\bibfnamefont
  {M.}~\bibnamefont {Greiner}}, \bibinfo {author} {\bibfnamefont
  {E.}~\bibnamefont {Demler}}, \bibinfo {author} {\bibfnamefont
  {F.}~\bibnamefont {Grusdt}}, \ and\ \bibinfo {author} {\bibfnamefont
  {M.}~\bibnamefont {Knap}},\ }\href
  {https://doi.org/10.1038/s41567-019-0565-x} {\bibfield  {journal} {\bibinfo
  {journal} {Nature Physics}\ }\textbf {\bibinfo {volume} {15}},\ \bibinfo
  {pages} {921} (\bibinfo {year} {2019})}\BibitemShut {NoStop}%
\bibitem [{\citenamefont {Chiu}\ \emph {et~al.}(2019)\citenamefont {Chiu},
  \citenamefont {Ji}, \citenamefont {Bohrdt}, \citenamefont {Xu}, \citenamefont
  {Knap}, \citenamefont {Demler}, \citenamefont {Grusdt}, \citenamefont
  {Greiner},\ and\ \citenamefont {Greif}}]{Chiu_Science_19_cold-atom}%
  \BibitemOpen
  \bibfield  {author} {\bibinfo {author} {\bibfnamefont {C.~S.}\ \bibnamefont
  {Chiu}}, \bibinfo {author} {\bibfnamefont {G.}~\bibnamefont {Ji}}, \bibinfo
  {author} {\bibfnamefont {A.}~\bibnamefont {Bohrdt}}, \bibinfo {author}
  {\bibfnamefont {M.}~\bibnamefont {Xu}}, \bibinfo {author} {\bibfnamefont
  {M.}~\bibnamefont {Knap}}, \bibinfo {author} {\bibfnamefont {E.}~\bibnamefont
  {Demler}}, \bibinfo {author} {\bibfnamefont {F.}~\bibnamefont {Grusdt}},
  \bibinfo {author} {\bibfnamefont {M.}~\bibnamefont {Greiner}}, \ and\
  \bibinfo {author} {\bibfnamefont {D.}~\bibnamefont {Greif}},\ }\href
  {\doibase 10.1126/science.aav3587} {\bibfield  {journal} {\bibinfo  {journal}
  {Science}\ }\textbf {\bibinfo {volume} {365}},\ \bibinfo {pages} {251}
  (\bibinfo {year} {2019})}\BibitemShut {NoStop}%
\bibitem [{\citenamefont {Brown}\ \emph {et~al.}(2019)\citenamefont {Brown},
  \citenamefont {Mitra}, \citenamefont {Guardado-Sanchez}, \citenamefont
  {Nourafkan}, \citenamefont {Reymbaut}, \citenamefont {Hébert}, \citenamefont
  {Bergeron}, \citenamefont {Tremblay}, \citenamefont {Kokalj}, \citenamefont
  {Huse}, \citenamefont {Schauß},\ and\ \citenamefont
  {Bakr}}]{Brown_Science_19_cold-atom}%
  \BibitemOpen
  \bibfield  {author} {\bibinfo {author} {\bibfnamefont {P.~T.}\ \bibnamefont
  {Brown}}, \bibinfo {author} {\bibfnamefont {D.}~\bibnamefont {Mitra}},
  \bibinfo {author} {\bibfnamefont {E.}~\bibnamefont {Guardado-Sanchez}},
  \bibinfo {author} {\bibfnamefont {R.}~\bibnamefont {Nourafkan}}, \bibinfo
  {author} {\bibfnamefont {A.}~\bibnamefont {Reymbaut}}, \bibinfo {author}
  {\bibfnamefont {C.-D.}\ \bibnamefont {Hébert}}, \bibinfo {author}
  {\bibfnamefont {S.}~\bibnamefont {Bergeron}}, \bibinfo {author}
  {\bibfnamefont {A.-M.~S.}\ \bibnamefont {Tremblay}}, \bibinfo {author}
  {\bibfnamefont {J.}~\bibnamefont {Kokalj}}, \bibinfo {author} {\bibfnamefont
  {D.~A.}\ \bibnamefont {Huse}}, \bibinfo {author} {\bibfnamefont
  {P.}~\bibnamefont {Schauß}}, \ and\ \bibinfo {author} {\bibfnamefont
  {W.~S.}\ \bibnamefont {Bakr}},\ }\href {\doibase 10.1126/science.aat4134}
  {\bibfield  {journal} {\bibinfo  {journal} {Science}\ }\textbf {\bibinfo
  {volume} {363}},\ \bibinfo {pages} {379} (\bibinfo {year}
  {2019})}\BibitemShut {NoStop}%
\bibitem [{\citenamefont {Mazurenko}\ \emph {et~al.}(2017)\citenamefont
  {Mazurenko}, \citenamefont {Chiu}, \citenamefont {Ji}, \citenamefont
  {Parsons}, \citenamefont {Kanász-Nagy}, \citenamefont {Schmidt},
  \citenamefont {Grusdt}, \citenamefont {Demler}, \citenamefont {Greif},\ and\
  \citenamefont {Greiner}}]{Mazurenko_Nature_17_cold-atom}%
  \BibitemOpen
  \bibfield  {author} {\bibinfo {author} {\bibfnamefont {A.}~\bibnamefont
  {Mazurenko}}, \bibinfo {author} {\bibfnamefont {C.~S.}\ \bibnamefont {Chiu}},
  \bibinfo {author} {\bibfnamefont {G.}~\bibnamefont {Ji}}, \bibinfo {author}
  {\bibfnamefont {M.~F.}\ \bibnamefont {Parsons}}, \bibinfo {author}
  {\bibfnamefont {M.}~\bibnamefont {Kanász-Nagy}}, \bibinfo {author}
  {\bibfnamefont {R.}~\bibnamefont {Schmidt}}, \bibinfo {author} {\bibfnamefont
  {F.}~\bibnamefont {Grusdt}}, \bibinfo {author} {\bibfnamefont
  {E.}~\bibnamefont {Demler}}, \bibinfo {author} {\bibfnamefont
  {D.}~\bibnamefont {Greif}}, \ and\ \bibinfo {author} {\bibfnamefont
  {M.}~\bibnamefont {Greiner}},\ }\href {https://doi.org/10.1038/nature22362}
  {\bibfield  {journal} {\bibinfo  {journal} {Nature}\ }\textbf {\bibinfo
  {volume} {545}},\ \bibinfo {pages} {462} (\bibinfo {year}
  {2017})}\BibitemShut {NoStop}%
\bibitem [{\citenamefont {Nyhegn}\ \emph {et~al.}(2022)\citenamefont {Nyhegn},
  \citenamefont {Nielsen},\ and\ \citenamefont
  {Bruun}}]{Nyhegn_PRB_22_bilayer-t-J_SCBA_dynamics}%
  \BibitemOpen
  \bibfield  {author} {\bibinfo {author} {\bibfnamefont {J.~H.}\ \bibnamefont
  {Nyhegn}}, \bibinfo {author} {\bibfnamefont {K.~K.}\ \bibnamefont {Nielsen}},
  \ and\ \bibinfo {author} {\bibfnamefont {G.~M.}\ \bibnamefont {Bruun}},\
  }\href {\doibase 10.1103/PhysRevB.106.155160} {\bibfield  {journal} {\bibinfo
   {journal} {Phys. Rev. B}\ }\textbf {\bibinfo {volume} {106}},\ \bibinfo
  {pages} {155160} (\bibinfo {year} {2022})}\BibitemShut {NoStop}%
\bibitem [{\citenamefont {Zhang}\ and\ \citenamefont
  {Rice}(1988)}]{Zhang-Rice_PRB_88}%
  \BibitemOpen
  \bibfield  {author} {\bibinfo {author} {\bibfnamefont {F.~C.}\ \bibnamefont
  {Zhang}}\ and\ \bibinfo {author} {\bibfnamefont {T.~M.}\ \bibnamefont
  {Rice}},\ }\href {\doibase 10.1103/PhysRevB.37.3759} {\bibfield  {journal}
  {\bibinfo  {journal} {Phys. Rev. B}\ }\textbf {\bibinfo {volume} {37}},\
  \bibinfo {pages} {3759} (\bibinfo {year} {1988})}\BibitemShut {NoStop}%
\bibitem [{\citenamefont {Damascelli}\ \emph {et~al.}(2003)\citenamefont
  {Damascelli}, \citenamefont {Hussain},\ and\ \citenamefont
  {Shen}}]{Damascelli_RMP_03_reivew_arpes}%
  \BibitemOpen
  \bibfield  {author} {\bibinfo {author} {\bibfnamefont {A.}~\bibnamefont
  {Damascelli}}, \bibinfo {author} {\bibfnamefont {Z.}~\bibnamefont {Hussain}},
  \ and\ \bibinfo {author} {\bibfnamefont {Z.-X.}\ \bibnamefont {Shen}},\
  }\href {\doibase 10.1103/RevModPhys.75.473} {\bibfield  {journal} {\bibinfo
  {journal} {Rev. Mod. Phys.}\ }\textbf {\bibinfo {volume} {75}},\ \bibinfo
  {pages} {473} (\bibinfo {year} {2003})}\BibitemShut {NoStop}%
\bibitem [{\citenamefont {Marshall}\ \emph {et~al.}(1996)\citenamefont
  {Marshall}, \citenamefont {Dessau}, \citenamefont {Loeser}, \citenamefont
  {Park}, \citenamefont {Matsuura}, \citenamefont {Eckstein}, \citenamefont
  {Bozovic}, \citenamefont {Fournier}, \citenamefont {Kapitulnik},
  \citenamefont {Spicer},\ and\ \citenamefont {Shen}}]{Marshall_PRL_96}%
  \BibitemOpen
  \bibfield  {author} {\bibinfo {author} {\bibfnamefont {D.~S.}\ \bibnamefont
  {Marshall}}, \bibinfo {author} {\bibfnamefont {D.~S.}\ \bibnamefont
  {Dessau}}, \bibinfo {author} {\bibfnamefont {A.~G.}\ \bibnamefont {Loeser}},
  \bibinfo {author} {\bibfnamefont {C.-H.}\ \bibnamefont {Park}}, \bibinfo
  {author} {\bibfnamefont {A.~Y.}\ \bibnamefont {Matsuura}}, \bibinfo {author}
  {\bibfnamefont {J.~N.}\ \bibnamefont {Eckstein}}, \bibinfo {author}
  {\bibfnamefont {I.}~\bibnamefont {Bozovic}}, \bibinfo {author} {\bibfnamefont
  {P.}~\bibnamefont {Fournier}}, \bibinfo {author} {\bibfnamefont
  {A.}~\bibnamefont {Kapitulnik}}, \bibinfo {author} {\bibfnamefont {W.~E.}\
  \bibnamefont {Spicer}}, \ and\ \bibinfo {author} {\bibfnamefont {Z.-X.}\
  \bibnamefont {Shen}},\ }\href {\doibase 10.1103/PhysRevLett.76.4841}
  {\bibfield  {journal} {\bibinfo  {journal} {Phys. Rev. Lett.}\ }\textbf
  {\bibinfo {volume} {76}},\ \bibinfo {pages} {4841} (\bibinfo {year}
  {1996})}\BibitemShut {NoStop}%
\bibitem [{\citenamefont {Eder}\ \emph {et~al.}(1997)\citenamefont {Eder},
  \citenamefont {Ohta},\ and\ \citenamefont {Sawatzky}}]{Eder_PRB_97}%
  \BibitemOpen
  \bibfield  {author} {\bibinfo {author} {\bibfnamefont {R.}~\bibnamefont
  {Eder}}, \bibinfo {author} {\bibfnamefont {Y.}~\bibnamefont {Ohta}}, \ and\
  \bibinfo {author} {\bibfnamefont {G.~A.}\ \bibnamefont {Sawatzky}},\ }\href
  {\doibase 10.1103/PhysRevB.55.R3414} {\bibfield  {journal} {\bibinfo
  {journal} {Phys. Rev. B}\ }\textbf {\bibinfo {volume} {55}},\ \bibinfo
  {pages} {R3414} (\bibinfo {year} {1997})}\BibitemShut {NoStop}%
\bibitem [{\citenamefont {Kim}\ \emph {et~al.}(1998)\citenamefont {Kim},
  \citenamefont {White}, \citenamefont {Shen}, \citenamefont {Tohyama},
  \citenamefont {Shibata}, \citenamefont {Maekawa}, \citenamefont {Wells},
  \citenamefont {Kim}, \citenamefont {Birgeneau},\ and\ \citenamefont
  {Kastner}}]{Kim_PRL_98}%
  \BibitemOpen
  \bibfield  {author} {\bibinfo {author} {\bibfnamefont {C.}~\bibnamefont
  {Kim}}, \bibinfo {author} {\bibfnamefont {P.~J.}\ \bibnamefont {White}},
  \bibinfo {author} {\bibfnamefont {Z.-X.}\ \bibnamefont {Shen}}, \bibinfo
  {author} {\bibfnamefont {T.}~\bibnamefont {Tohyama}}, \bibinfo {author}
  {\bibfnamefont {Y.}~\bibnamefont {Shibata}}, \bibinfo {author} {\bibfnamefont
  {S.}~\bibnamefont {Maekawa}}, \bibinfo {author} {\bibfnamefont {B.~O.}\
  \bibnamefont {Wells}}, \bibinfo {author} {\bibfnamefont {Y.~J.}\ \bibnamefont
  {Kim}}, \bibinfo {author} {\bibfnamefont {R.~J.}\ \bibnamefont {Birgeneau}},
  \ and\ \bibinfo {author} {\bibfnamefont {M.~A.}\ \bibnamefont {Kastner}},\
  }\href {\doibase 10.1103/PhysRevLett.80.4245} {\bibfield  {journal} {\bibinfo
   {journal} {Phys. Rev. Lett.}\ }\textbf {\bibinfo {volume} {80}},\ \bibinfo
  {pages} {4245} (\bibinfo {year} {1998})}\BibitemShut {NoStop}%
\bibitem [{\citenamefont {Yin}\ \emph {et~al.}(1998)\citenamefont {Yin},
  \citenamefont {Gong},\ and\ \citenamefont {Leung}}]{Yin_PRL_98}%
  \BibitemOpen
  \bibfield  {author} {\bibinfo {author} {\bibfnamefont {W.-G.}\ \bibnamefont
  {Yin}}, \bibinfo {author} {\bibfnamefont {C.-D.}\ \bibnamefont {Gong}}, \
  and\ \bibinfo {author} {\bibfnamefont {P.~W.}\ \bibnamefont {Leung}},\ }\href
  {\doibase 10.1103/PhysRevLett.81.2534} {\bibfield  {journal} {\bibinfo
  {journal} {Phys. Rev. Lett.}\ }\textbf {\bibinfo {volume} {81}},\ \bibinfo
  {pages} {2534} (\bibinfo {year} {1998})}\BibitemShut {NoStop}%
\bibitem [{\citenamefont {Wells}\ \emph {et~al.}(1995)\citenamefont {Wells},
  \citenamefont {Shen}, \citenamefont {Matsuura}, \citenamefont {King},
  \citenamefont {Kastner}, \citenamefont {Greven},\ and\ \citenamefont
  {Birgeneau}}]{Wells_PRL_95_Sr2CuO2Cl2}%
  \BibitemOpen
  \bibfield  {author} {\bibinfo {author} {\bibfnamefont {B.~O.}\ \bibnamefont
  {Wells}}, \bibinfo {author} {\bibfnamefont {Z.~X.}\ \bibnamefont {Shen}},
  \bibinfo {author} {\bibfnamefont {A.}~\bibnamefont {Matsuura}}, \bibinfo
  {author} {\bibfnamefont {D.~M.}\ \bibnamefont {King}}, \bibinfo {author}
  {\bibfnamefont {M.~A.}\ \bibnamefont {Kastner}}, \bibinfo {author}
  {\bibfnamefont {M.}~\bibnamefont {Greven}}, \ and\ \bibinfo {author}
  {\bibfnamefont {R.~J.}\ \bibnamefont {Birgeneau}},\ }\href {\doibase
  10.1103/PhysRevLett.74.964} {\bibfield  {journal} {\bibinfo  {journal} {Phys.
  Rev. Lett.}\ }\textbf {\bibinfo {volume} {74}},\ \bibinfo {pages} {964}
  (\bibinfo {year} {1995})}\BibitemShut {NoStop}%
\bibitem [{\citenamefont {Ronning}\ \emph {et~al.}(2003)\citenamefont
  {Ronning}, \citenamefont {Kim}, \citenamefont {Shen}, \citenamefont
  {Armitage}, \citenamefont {Damascelli}, \citenamefont {Lu}, \citenamefont
  {Feng}, \citenamefont {Shen}, \citenamefont {Miller}, \citenamefont {Kim},
  \citenamefont {Chou},\ and\ \citenamefont
  {Terasaki}}]{Ronning_PRB_03_undoped_arpes}%
  \BibitemOpen
  \bibfield  {author} {\bibinfo {author} {\bibfnamefont {F.}~\bibnamefont
  {Ronning}}, \bibinfo {author} {\bibfnamefont {C.}~\bibnamefont {Kim}},
  \bibinfo {author} {\bibfnamefont {K.~M.}\ \bibnamefont {Shen}}, \bibinfo
  {author} {\bibfnamefont {N.~P.}\ \bibnamefont {Armitage}}, \bibinfo {author}
  {\bibfnamefont {A.}~\bibnamefont {Damascelli}}, \bibinfo {author}
  {\bibfnamefont {D.~H.}\ \bibnamefont {Lu}}, \bibinfo {author} {\bibfnamefont
  {D.~L.}\ \bibnamefont {Feng}}, \bibinfo {author} {\bibfnamefont {Z.-X.}\
  \bibnamefont {Shen}}, \bibinfo {author} {\bibfnamefont {L.~L.}\ \bibnamefont
  {Miller}}, \bibinfo {author} {\bibfnamefont {Y.-J.}\ \bibnamefont {Kim}},
  \bibinfo {author} {\bibfnamefont {F.}~\bibnamefont {Chou}}, \ and\ \bibinfo
  {author} {\bibfnamefont {I.}~\bibnamefont {Terasaki}},\ }\href {\doibase
  10.1103/PhysRevB.67.035113} {\bibfield  {journal} {\bibinfo  {journal} {Phys.
  Rev. B}\ }\textbf {\bibinfo {volume} {67}},\ \bibinfo {pages} {035113}
  (\bibinfo {year} {2003})}\BibitemShut {NoStop}%
\bibitem [{\citenamefont {Nazarenko}\ \emph {et~al.}(1995)\citenamefont
  {Nazarenko}, \citenamefont {Vos}, \citenamefont {Haas}, \citenamefont
  {Dagotto},\ and\ \citenamefont {Gooding}}]{Nazarenko_PRB_95_Sr2CuO2Cl2}%
  \BibitemOpen
  \bibfield  {author} {\bibinfo {author} {\bibfnamefont {A.}~\bibnamefont
  {Nazarenko}}, \bibinfo {author} {\bibfnamefont {K.~J.~E.}\ \bibnamefont
  {Vos}}, \bibinfo {author} {\bibfnamefont {S.}~\bibnamefont {Haas}}, \bibinfo
  {author} {\bibfnamefont {E.}~\bibnamefont {Dagotto}}, \ and\ \bibinfo
  {author} {\bibfnamefont {R.~J.}\ \bibnamefont {Gooding}},\ }\href {\doibase
  10.1103/PhysRevB.51.8676} {\bibfield  {journal} {\bibinfo  {journal} {Phys.
  Rev. B}\ }\textbf {\bibinfo {volume} {51}},\ \bibinfo {pages} {8676}
  (\bibinfo {year} {1995})}\BibitemShut {NoStop}%
\bibitem [{\citenamefont {Kyung}\ and\ \citenamefont
  {Ferrell}(1996)}]{Kyung_PRB_96_Sr2CuO2Cl2}%
  \BibitemOpen
  \bibfield  {author} {\bibinfo {author} {\bibfnamefont {B.}~\bibnamefont
  {Kyung}}\ and\ \bibinfo {author} {\bibfnamefont {R.~A.}\ \bibnamefont
  {Ferrell}},\ }\href {\doibase 10.1103/PhysRevB.54.10125} {\bibfield
  {journal} {\bibinfo  {journal} {Phys. Rev. B}\ }\textbf {\bibinfo {volume}
  {54}},\ \bibinfo {pages} {10125} (\bibinfo {year} {1996})}\BibitemShut
  {NoStop}%
\bibitem [{\citenamefont {Xiang}\ and\ \citenamefont
  {Wheatley}(1996)}]{Xiang_PRB_96_Sr2CuO2Cl2}%
  \BibitemOpen
  \bibfield  {author} {\bibinfo {author} {\bibfnamefont {T.}~\bibnamefont
  {Xiang}}\ and\ \bibinfo {author} {\bibfnamefont {J.~M.}\ \bibnamefont
  {Wheatley}},\ }\href {\doibase 10.1103/PhysRevB.54.R12653} {\bibfield
  {journal} {\bibinfo  {journal} {Phys. Rev. B}\ }\textbf {\bibinfo {volume}
  {54}},\ \bibinfo {pages} {R12653} (\bibinfo {year} {1996})}\BibitemShut
  {NoStop}%
\bibitem [{\citenamefont {Yin}\ and\ \citenamefont
  {Ku}(2009{\natexlab{a}})}]{Yin_PRB_09_WF}%
  \BibitemOpen
  \bibfield  {author} {\bibinfo {author} {\bibfnamefont {W.-G.}\ \bibnamefont
  {Yin}}\ and\ \bibinfo {author} {\bibfnamefont {W.}~\bibnamefont {Ku}},\
  }\href {\doibase 10.1103/PhysRevB.79.214512} {\bibfield  {journal} {\bibinfo
  {journal} {Phys. Rev. B}\ }\textbf {\bibinfo {volume} {79}},\ \bibinfo
  {pages} {214512} (\bibinfo {year} {2009}{\natexlab{a}})}\BibitemShut
  {NoStop}%
\bibitem [{\citenamefont {Belinicher}\ \emph {et~al.}(1996)\citenamefont
  {Belinicher}, \citenamefont {Chernyshev},\ and\ \citenamefont
  {Shubin}}]{Belinicher_PRB_96_Sr2CuO2Cl2}%
  \BibitemOpen
  \bibfield  {author} {\bibinfo {author} {\bibfnamefont {V.~I.}\ \bibnamefont
  {Belinicher}}, \bibinfo {author} {\bibfnamefont {A.~L.}\ \bibnamefont
  {Chernyshev}}, \ and\ \bibinfo {author} {\bibfnamefont {V.~A.}\ \bibnamefont
  {Shubin}},\ }\href {\doibase 10.1103/PhysRevB.54.14914} {\bibfield  {journal}
  {\bibinfo  {journal} {Phys. Rev. B}\ }\textbf {\bibinfo {volume} {54}},\
  \bibinfo {pages} {14914} (\bibinfo {year} {1996})}\BibitemShut {NoStop}%
\bibitem [{\citenamefont {Leung}\ \emph {et~al.}(1997)\citenamefont {Leung},
  \citenamefont {Wells},\ and\ \citenamefont
  {Gooding}}]{Leung_PRB_97_Sr2CuO2Cl2}%
  \BibitemOpen
  \bibfield  {author} {\bibinfo {author} {\bibfnamefont {P.~W.}\ \bibnamefont
  {Leung}}, \bibinfo {author} {\bibfnamefont {B.~O.}\ \bibnamefont {Wells}}, \
  and\ \bibinfo {author} {\bibfnamefont {R.~J.}\ \bibnamefont {Gooding}},\
  }\href {\doibase 10.1103/PhysRevB.56.6320} {\bibfield  {journal} {\bibinfo
  {journal} {Phys. Rev. B}\ }\textbf {\bibinfo {volume} {56}},\ \bibinfo
  {pages} {6320} (\bibinfo {year} {1997})}\BibitemShut {NoStop}%
\bibitem [{\citenamefont {Lee}\ and\ \citenamefont
  {Shih}(1997)}]{Lee_PRB_97_Sr2CuO2Cl2}%
  \BibitemOpen
  \bibfield  {author} {\bibinfo {author} {\bibfnamefont {T.~K.}\ \bibnamefont
  {Lee}}\ and\ \bibinfo {author} {\bibfnamefont {C.~T.}\ \bibnamefont {Shih}},\
  }\href {\doibase 10.1103/PhysRevB.55.5983} {\bibfield  {journal} {\bibinfo
  {journal} {Phys. Rev. B}\ }\textbf {\bibinfo {volume} {55}},\ \bibinfo
  {pages} {5983} (\bibinfo {year} {1997})}\BibitemShut {NoStop}%
\bibitem [{\citenamefont {Pavarini}\ \emph {et~al.}(2001)\citenamefont
  {Pavarini}, \citenamefont {Dasgupta}, \citenamefont {Saha-Dasgupta},
  \citenamefont {Jepsen},\ and\ \citenamefont {Andersen}}]{Pavarini_PRL_01_WF}%
  \BibitemOpen
  \bibfield  {author} {\bibinfo {author} {\bibfnamefont {E.}~\bibnamefont
  {Pavarini}}, \bibinfo {author} {\bibfnamefont {I.}~\bibnamefont {Dasgupta}},
  \bibinfo {author} {\bibfnamefont {T.}~\bibnamefont {Saha-Dasgupta}}, \bibinfo
  {author} {\bibfnamefont {O.}~\bibnamefont {Jepsen}}, \ and\ \bibinfo {author}
  {\bibfnamefont {O.~K.}\ \bibnamefont {Andersen}},\ }\href {\doibase
  10.1103/PhysRevLett.87.047003} {\bibfield  {journal} {\bibinfo  {journal}
  {Phys. Rev. Lett.}\ }\textbf {\bibinfo {volume} {87}},\ \bibinfo {pages}
  {047003} (\bibinfo {year} {2001})}\BibitemShut {NoStop}%
\bibitem [{\citenamefont {Yu}\ \emph {et~al.}(2017)\citenamefont {Yu},
  \citenamefont {Zhou}, \citenamefont {Yin}, \citenamefont {Lin},\ and\
  \citenamefont {Gong}}]{Yin_PRB_17_phase-competition}%
  \BibitemOpen
  \bibfield  {author} {\bibinfo {author} {\bibfnamefont {Z.-D.}\ \bibnamefont
  {Yu}}, \bibinfo {author} {\bibfnamefont {Y.}~\bibnamefont {Zhou}}, \bibinfo
  {author} {\bibfnamefont {W.-G.}\ \bibnamefont {Yin}}, \bibinfo {author}
  {\bibfnamefont {H.-Q.}\ \bibnamefont {Lin}}, \ and\ \bibinfo {author}
  {\bibfnamefont {C.-D.}\ \bibnamefont {Gong}},\ }\href {\doibase
  10.1103/PhysRevB.96.045110} {\bibfield  {journal} {\bibinfo  {journal} {Phys.
  Rev. B}\ }\textbf {\bibinfo {volume} {96}},\ \bibinfo {pages} {045110}
  (\bibinfo {year} {2017})}\BibitemShut {NoStop}%
\bibitem [{\citenamefont {VALLA}\ \emph {et~al.}(1999)\citenamefont {VALLA},
  \citenamefont {FEDOROV}, \citenamefont {JOHNSON}, \citenamefont {WELLS},
  \citenamefont {HULBERT}, \citenamefont {LI}, \citenamefont {GU},\ and\
  \citenamefont {KOSHIZUKA}}]{Valla_Science_99_MDC}%
  \BibitemOpen
  \bibfield  {author} {\bibinfo {author} {\bibfnamefont {T.}~\bibnamefont
  {VALLA}}, \bibinfo {author} {\bibfnamefont {A.~V.}\ \bibnamefont {FEDOROV}},
  \bibinfo {author} {\bibfnamefont {P.~D.}\ \bibnamefont {JOHNSON}}, \bibinfo
  {author} {\bibfnamefont {B.~O.}\ \bibnamefont {WELLS}}, \bibinfo {author}
  {\bibfnamefont {S.~L.}\ \bibnamefont {HULBERT}}, \bibinfo {author}
  {\bibfnamefont {Q.}~\bibnamefont {LI}}, \bibinfo {author} {\bibfnamefont
  {G.~D.}\ \bibnamefont {GU}}, \ and\ \bibinfo {author} {\bibfnamefont
  {N.}~\bibnamefont {KOSHIZUKA}},\ }\href {\doibase
  10.1126/science.285.5436.2110} {\bibfield  {journal} {\bibinfo  {journal}
  {Science}\ }\textbf {\bibinfo {volume} {285}},\ \bibinfo {pages} {2110}
  (\bibinfo {year} {1999})}\BibitemShut {NoStop}%
\bibitem [{\citenamefont
  {Laughlin}(1997)}]{Laughlin_PRL_97_Sr2CuO2Cl2_slave-boson}%
  \BibitemOpen
  \bibfield  {author} {\bibinfo {author} {\bibfnamefont {R.~B.}\ \bibnamefont
  {Laughlin}},\ }\href {\doibase 10.1103/PhysRevLett.79.1726} {\bibfield
  {journal} {\bibinfo  {journal} {Phys. Rev. Lett.}\ }\textbf {\bibinfo
  {volume} {79}},\ \bibinfo {pages} {1726} (\bibinfo {year}
  {1997})}\BibitemShut {NoStop}%
\bibitem [{\citenamefont {Krenn}\ \emph {et~al.}(2022)\citenamefont {Krenn},
  \citenamefont {Pollice}, \citenamefont {Guo}, \citenamefont {Aldeghi},
  \citenamefont {Cervera-Lierta}, \citenamefont {Friederich}, \citenamefont
  {dos Passos Gomes}, \citenamefont {Häse}, \citenamefont {Jinich},
  \citenamefont {Nigam}, \citenamefont {Yao},\ and\ \citenamefont
  {Aspuru-Guzik}}]{Krenn_NRP_22_ML_review}%
  \BibitemOpen
  \bibfield  {author} {\bibinfo {author} {\bibfnamefont {M.}~\bibnamefont
  {Krenn}}, \bibinfo {author} {\bibfnamefont {R.}~\bibnamefont {Pollice}},
  \bibinfo {author} {\bibfnamefont {S.~Y.}\ \bibnamefont {Guo}}, \bibinfo
  {author} {\bibfnamefont {M.}~\bibnamefont {Aldeghi}}, \bibinfo {author}
  {\bibfnamefont {A.}~\bibnamefont {Cervera-Lierta}}, \bibinfo {author}
  {\bibfnamefont {P.}~\bibnamefont {Friederich}}, \bibinfo {author}
  {\bibfnamefont {G.}~\bibnamefont {dos Passos Gomes}}, \bibinfo {author}
  {\bibfnamefont {F.}~\bibnamefont {Häse}}, \bibinfo {author} {\bibfnamefont
  {A.}~\bibnamefont {Jinich}}, \bibinfo {author} {\bibfnamefont
  {A.}~\bibnamefont {Nigam}}, \bibinfo {author} {\bibfnamefont
  {Z.}~\bibnamefont {Yao}}, \ and\ \bibinfo {author} {\bibfnamefont
  {A.}~\bibnamefont {Aspuru-Guzik}},\ }\href {\doibase
  10.1038/s42254-022-00518-3} {\bibfield  {journal} {\bibinfo  {journal}
  {Nature Reviews Physics}\ }\textbf {\bibinfo {volume} {4}},\ \bibinfo {pages}
  {761} (\bibinfo {year} {2022})}\BibitemShut {NoStop}%
\bibitem [{\citenamefont {Radovic}\ \emph {et~al.}(2018)\citenamefont
  {Radovic}, \citenamefont {Williams}, \citenamefont {Rousseau}, \citenamefont
  {Kagan}, \citenamefont {Bonacorsi}, \citenamefont {Himmel}, \citenamefont
  {Aurisano}, \citenamefont {Terao},\ and\ \citenamefont
  {Wongjirad}}]{Radovic_Nature_18_ML_ParticlePhys}%
  \BibitemOpen
  \bibfield  {author} {\bibinfo {author} {\bibfnamefont {A.}~\bibnamefont
  {Radovic}}, \bibinfo {author} {\bibfnamefont {M.}~\bibnamefont {Williams}},
  \bibinfo {author} {\bibfnamefont {D.}~\bibnamefont {Rousseau}}, \bibinfo
  {author} {\bibfnamefont {M.}~\bibnamefont {Kagan}}, \bibinfo {author}
  {\bibfnamefont {D.}~\bibnamefont {Bonacorsi}}, \bibinfo {author}
  {\bibfnamefont {A.}~\bibnamefont {Himmel}}, \bibinfo {author} {\bibfnamefont
  {A.}~\bibnamefont {Aurisano}}, \bibinfo {author} {\bibfnamefont
  {K.}~\bibnamefont {Terao}}, \ and\ \bibinfo {author} {\bibfnamefont
  {T.}~\bibnamefont {Wongjirad}},\ }\href
  {https://doi.org/10.1038/s41586-018-0361-2} {\bibfield  {journal} {\bibinfo
  {journal} {Nature}\ }\textbf {\bibinfo {volume} {560}},\ \bibinfo {pages}
  {41} (\bibinfo {year} {2018})}\BibitemShut {NoStop}%
\bibitem [{\citenamefont {Sadeh}\ \emph {et~al.}(2016)\citenamefont {Sadeh},
  \citenamefont {Abdalla},\ and\ \citenamefont {Lahav}}]{Sadeh_16_ML_redshift}%
  \BibitemOpen
  \bibfield  {author} {\bibinfo {author} {\bibfnamefont {I.}~\bibnamefont
  {Sadeh}}, \bibinfo {author} {\bibfnamefont {F.~B.}\ \bibnamefont {Abdalla}},
  \ and\ \bibinfo {author} {\bibfnamefont {O.}~\bibnamefont {Lahav}},\ }\href
  {\doibase 10.1088/1538-3873/128/968/104502} {\bibfield  {journal} {\bibinfo
  {journal} {Publications of the Astronomical Society of the Pacific}\ }\textbf
  {\bibinfo {volume} {128}},\ \bibinfo {pages} {104502} (\bibinfo {year}
  {2016})}\BibitemShut {NoStop}%
\bibitem [{\citenamefont {Hashimoto}\ and\ \citenamefont
  {Liu}(2022)}]{Hashimoto_Universe_22_ML_galaxy}%
  \BibitemOpen
  \bibfield  {author} {\bibinfo {author} {\bibfnamefont {Y.}~\bibnamefont
  {Hashimoto}}\ and\ \bibinfo {author} {\bibfnamefont {C.-H.}\ \bibnamefont
  {Liu}},\ }\href {\doibase 10.3390/universe8070339} {\bibfield  {journal}
  {\bibinfo  {journal} {Universe}\ }\textbf {\bibinfo {volume} {8}},\ \bibinfo
  {pages} {339} (\bibinfo {year} {2022})}\BibitemShut {NoStop}%
\bibitem [{\citenamefont {Schanche}\ \emph {et~al.}(2018)\citenamefont
  {Schanche}, \citenamefont {Cameron}, \citenamefont {Hébrard}, \citenamefont
  {Nielsen}, \citenamefont {Triaud}, \citenamefont {Almenara}, \citenamefont
  {Alsubai}, \citenamefont {Anderson}, \citenamefont {Armstrong}, \citenamefont
  {Barros}, \citenamefont {Bouchy}, \citenamefont {Boumis}, \citenamefont
  {Brown}, \citenamefont {Faedi}, \citenamefont {Hay}, \citenamefont {Hebb},
  \citenamefont {Kiefer}, \citenamefont {Mancini}, \citenamefont {Maxted},
  \citenamefont {Palle}, \citenamefont {Pollacco}, \citenamefont {Queloz},
  \citenamefont {Smalley}, \citenamefont {Udry}, \citenamefont {West},\ and\
  \citenamefont {Wheatley}}]{Schanche_19_ML_ExoplanetTransit}%
  \BibitemOpen
  \bibfield  {author} {\bibinfo {author} {\bibfnamefont {N.}~\bibnamefont
  {Schanche}}, \bibinfo {author} {\bibfnamefont {A.~C.}\ \bibnamefont
  {Cameron}}, \bibinfo {author} {\bibfnamefont {G.}~\bibnamefont {Hébrard}},
  \bibinfo {author} {\bibfnamefont {L.}~\bibnamefont {Nielsen}}, \bibinfo
  {author} {\bibfnamefont {A.~H. M.~J.}\ \bibnamefont {Triaud}}, \bibinfo
  {author} {\bibfnamefont {J.~M.}\ \bibnamefont {Almenara}}, \bibinfo {author}
  {\bibfnamefont {K.~A.}\ \bibnamefont {Alsubai}}, \bibinfo {author}
  {\bibfnamefont {D.~R.}\ \bibnamefont {Anderson}}, \bibinfo {author}
  {\bibfnamefont {D.~J.}\ \bibnamefont {Armstrong}}, \bibinfo {author}
  {\bibfnamefont {S.~C.~C.}\ \bibnamefont {Barros}}, \bibinfo {author}
  {\bibfnamefont {F.}~\bibnamefont {Bouchy}}, \bibinfo {author} {\bibfnamefont
  {P.}~\bibnamefont {Boumis}}, \bibinfo {author} {\bibfnamefont {D.~J.~A.}\
  \bibnamefont {Brown}}, \bibinfo {author} {\bibfnamefont {F.}~\bibnamefont
  {Faedi}}, \bibinfo {author} {\bibfnamefont {K.}~\bibnamefont {Hay}}, \bibinfo
  {author} {\bibfnamefont {L.}~\bibnamefont {Hebb}}, \bibinfo {author}
  {\bibfnamefont {F.}~\bibnamefont {Kiefer}}, \bibinfo {author} {\bibfnamefont
  {L.}~\bibnamefont {Mancini}}, \bibinfo {author} {\bibfnamefont {P.~F.~L.}\
  \bibnamefont {Maxted}}, \bibinfo {author} {\bibfnamefont {E.}~\bibnamefont
  {Palle}}, \bibinfo {author} {\bibfnamefont {D.~L.}\ \bibnamefont {Pollacco}},
  \bibinfo {author} {\bibfnamefont {D.}~\bibnamefont {Queloz}}, \bibinfo
  {author} {\bibfnamefont {B.}~\bibnamefont {Smalley}}, \bibinfo {author}
  {\bibfnamefont {S.}~\bibnamefont {Udry}}, \bibinfo {author} {\bibfnamefont
  {R.}~\bibnamefont {West}}, \ and\ \bibinfo {author} {\bibfnamefont {P.~J.}\
  \bibnamefont {Wheatley}},\ }\href {\doibase 10.1093/mnras/sty3146} {\bibfield
   {journal} {\bibinfo  {journal} {Monthly Notices of the Royal Astronomical
  Society}\ }\textbf {\bibinfo {volume} {483}},\ \bibinfo {pages} {5534}
  (\bibinfo {year} {2018})}\BibitemShut {NoStop}%
\bibitem [{\citenamefont {Schmidt}\ \emph {et~al.}(2019)\citenamefont
  {Schmidt}, \citenamefont {Marques}, \citenamefont {Botti},\ and\
  \citenamefont {Marques}}]{Schmidt_npjCM_19_ML_review}%
  \BibitemOpen
  \bibfield  {author} {\bibinfo {author} {\bibfnamefont {J.}~\bibnamefont
  {Schmidt}}, \bibinfo {author} {\bibfnamefont {M.~R.~G.}\ \bibnamefont
  {Marques}}, \bibinfo {author} {\bibfnamefont {S.}~\bibnamefont {Botti}}, \
  and\ \bibinfo {author} {\bibfnamefont {M.~A.~L.}\ \bibnamefont {Marques}},\
  }\href {\doibase 10.1038/s41524-019-0221-0} {\bibfield  {journal} {\bibinfo
  {journal} {npj Computational Materials}\ }\textbf {\bibinfo {volume} {5}},\
  \bibinfo {pages} {83} (\bibinfo {year} {2019})}\BibitemShut {NoStop}%
\bibitem [{\citenamefont {Ryan}\ \emph {et~al.}(2018)\citenamefont {Ryan},
  \citenamefont {Lengyel},\ and\ \citenamefont
  {Shatruk}}]{Ryan_18_JACS_ML_crystalStruct}%
  \BibitemOpen
  \bibfield  {author} {\bibinfo {author} {\bibfnamefont {K.}~\bibnamefont
  {Ryan}}, \bibinfo {author} {\bibfnamefont {J.}~\bibnamefont {Lengyel}}, \
  and\ \bibinfo {author} {\bibfnamefont {M.}~\bibnamefont {Shatruk}},\ }\href
  {\doibase 10.1021/jacs.8b03913} {\bibfield  {journal} {\bibinfo  {journal}
  {Journal of the American Chemical Society}\ }\textbf {\bibinfo {volume}
  {140}},\ \bibinfo {pages} {10158} (\bibinfo {year} {2018})}\BibitemShut
  {NoStop}%
\bibitem [{\citenamefont {Zheng}\ \emph {et~al.}(2018)\citenamefont {Zheng},
  \citenamefont {Zheng},\ and\ \citenamefont
  {Zhang}}]{Zheng_ChemSci_18_ML_properties}%
  \BibitemOpen
  \bibfield  {author} {\bibinfo {author} {\bibfnamefont {X.}~\bibnamefont
  {Zheng}}, \bibinfo {author} {\bibfnamefont {P.}~\bibnamefont {Zheng}}, \ and\
  \bibinfo {author} {\bibfnamefont {R.-Z.}\ \bibnamefont {Zhang}},\ }\href
  {\doibase 10.1039/C8SC02648C} {\bibfield  {journal} {\bibinfo  {journal}
  {Chem. Sci.}\ }\textbf {\bibinfo {volume} {9}},\ \bibinfo {pages} {8426}
  (\bibinfo {year} {2018})}\BibitemShut {NoStop}%
\bibitem [{\citenamefont {Carbone}\ \emph {et~al.}(2019)\citenamefont
  {Carbone}, \citenamefont {Yoo}, \citenamefont {Topsakal},\ and\ \citenamefont
  {Lu}}]{carbone2019classification}%
  \BibitemOpen
  \bibfield  {author} {\bibinfo {author} {\bibfnamefont {M.~R.}\ \bibnamefont
  {Carbone}}, \bibinfo {author} {\bibfnamefont {S.}~\bibnamefont {Yoo}},
  \bibinfo {author} {\bibfnamefont {M.}~\bibnamefont {Topsakal}}, \ and\
  \bibinfo {author} {\bibfnamefont {D.}~\bibnamefont {Lu}},\ }\href {\doibase
  10.1103/PhysRevMaterials.3.033604} {\bibfield  {journal} {\bibinfo  {journal}
  {Phys. Rev. Mater.}\ }\textbf {\bibinfo {volume} {3}},\ \bibinfo {pages}
  {033604} (\bibinfo {year} {2019})}\BibitemShut {NoStop}%
\bibitem [{\citenamefont {Torrisi}\ \emph {et~al.}(2020)\citenamefont
  {Torrisi}, \citenamefont {Carbone}, \citenamefont {Rohr}, \citenamefont
  {Montoya}, \citenamefont {Ha}, \citenamefont {Yano}, \citenamefont {Suram},\
  and\ \citenamefont {Hung}}]{torrisi2020random}%
  \BibitemOpen
  \bibfield  {author} {\bibinfo {author} {\bibfnamefont {S.~B.}\ \bibnamefont
  {Torrisi}}, \bibinfo {author} {\bibfnamefont {M.~R.}\ \bibnamefont
  {Carbone}}, \bibinfo {author} {\bibfnamefont {B.~A.}\ \bibnamefont {Rohr}},
  \bibinfo {author} {\bibfnamefont {J.~H.}\ \bibnamefont {Montoya}}, \bibinfo
  {author} {\bibfnamefont {Y.}~\bibnamefont {Ha}}, \bibinfo {author}
  {\bibfnamefont {J.}~\bibnamefont {Yano}}, \bibinfo {author} {\bibfnamefont
  {S.~K.}\ \bibnamefont {Suram}}, \ and\ \bibinfo {author} {\bibfnamefont
  {L.}~\bibnamefont {Hung}},\ }\href {\doibase 10.1038/s41524-020-00376-6}
  {\bibfield  {journal} {\bibinfo  {journal} {npj Comput. Mater.}\ }\textbf
  {\bibinfo {volume} {6}},\ \bibinfo {pages} {1} (\bibinfo {year}
  {2020})}\BibitemShut {NoStop}%
\bibitem [{\citenamefont {Jalem}\ \emph {et~al.}(2018)\citenamefont {Jalem},
  \citenamefont {Kanamori}, \citenamefont {Takeuchi}, \citenamefont {Nakayama},
  \citenamefont {Yamasaki},\ and\ \citenamefont
  {Saito}}]{Jalem_SR_18_ML_fastIonConduct}%
  \BibitemOpen
  \bibfield  {author} {\bibinfo {author} {\bibfnamefont {R.}~\bibnamefont
  {Jalem}}, \bibinfo {author} {\bibfnamefont {K.}~\bibnamefont {Kanamori}},
  \bibinfo {author} {\bibfnamefont {I.}~\bibnamefont {Takeuchi}}, \bibinfo
  {author} {\bibfnamefont {M.}~\bibnamefont {Nakayama}}, \bibinfo {author}
  {\bibfnamefont {H.}~\bibnamefont {Yamasaki}}, \ and\ \bibinfo {author}
  {\bibfnamefont {T.}~\bibnamefont {Saito}},\ }\href
  {https://doi.org/10.1038/s41598-018-23852-y} {\bibfield  {journal} {\bibinfo
  {journal} {Scientific Reports}\ }\textbf {\bibinfo {volume} {8}},\ \bibinfo
  {pages} {5845} (\bibinfo {year} {2018})}\BibitemShut {NoStop}%
\bibitem [{\citenamefont {Carbone}\ \emph {et~al.}(2020)\citenamefont
  {Carbone}, \citenamefont {Topsakal}, \citenamefont {Lu},\ and\ \citenamefont
  {Yoo}}]{carbone2020machine}%
  \BibitemOpen
  \bibfield  {author} {\bibinfo {author} {\bibfnamefont {M.~R.}\ \bibnamefont
  {Carbone}}, \bibinfo {author} {\bibfnamefont {M.}~\bibnamefont {Topsakal}},
  \bibinfo {author} {\bibfnamefont {D.}~\bibnamefont {Lu}}, \ and\ \bibinfo
  {author} {\bibfnamefont {S.}~\bibnamefont {Yoo}},\ }\href {\doibase
  10.1103/PhysRevLett.124.156401} {\bibfield  {journal} {\bibinfo  {journal}
  {Phys. Rev. Lett.}\ }\textbf {\bibinfo {volume} {124}},\ \bibinfo {pages}
  {156401} (\bibinfo {year} {2020})}\BibitemShut {NoStop}%
\bibitem [{\citenamefont {Ghose}\ \emph {et~al.}(2022)\citenamefont {Ghose},
  \citenamefont {Segal}, \citenamefont {Meng}, \citenamefont {Liang},
  \citenamefont {Hybertsen}, \citenamefont {Qu}, \citenamefont {Stavitski},
  \citenamefont {Yoo}, \citenamefont {Lu},\ and\ \citenamefont
  {Carbone}}]{ghose2022uncertainty}%
  \BibitemOpen
  \bibfield  {author} {\bibinfo {author} {\bibfnamefont {A.}~\bibnamefont
  {Ghose}}, \bibinfo {author} {\bibfnamefont {M.}~\bibnamefont {Segal}},
  \bibinfo {author} {\bibfnamefont {F.}~\bibnamefont {Meng}}, \bibinfo {author}
  {\bibfnamefont {Z.}~\bibnamefont {Liang}}, \bibinfo {author} {\bibfnamefont
  {M.~S.}\ \bibnamefont {Hybertsen}}, \bibinfo {author} {\bibfnamefont
  {X.}~\bibnamefont {Qu}}, \bibinfo {author} {\bibfnamefont {E.}~\bibnamefont
  {Stavitski}}, \bibinfo {author} {\bibfnamefont {S.}~\bibnamefont {Yoo}},
  \bibinfo {author} {\bibfnamefont {D.}~\bibnamefont {Lu}}, \ and\ \bibinfo
  {author} {\bibfnamefont {M.~R.}\ \bibnamefont {Carbone}},\ }\href
  {https://doi.org/10.48550/arXiv.2210.00336} {\  (\bibinfo {year} {2022})},\
  \Eprint {http://arxiv.org/abs/2210.00336} {arXiv:2210.00336} \BibitemShut
  {NoStop}%
\bibitem [{\citenamefont {Rankine}\ and\ \citenamefont
  {Penfold}(2022)}]{rankine2022accurate}%
  \BibitemOpen
  \bibfield  {author} {\bibinfo {author} {\bibfnamefont {C.~D.}\ \bibnamefont
  {Rankine}}\ and\ \bibinfo {author} {\bibfnamefont {T.}~\bibnamefont
  {Penfold}},\ }\href {\doibase 10.1063/5.0087255} {\bibfield  {journal}
  {\bibinfo  {journal} {J. Chem. Phys.}\ }\textbf {\bibinfo {volume} {156}},\
  \bibinfo {pages} {164102} (\bibinfo {year} {2022})}\BibitemShut {NoStop}%
\bibitem [{\citenamefont {Penfold}\ and\ \citenamefont
  {Rankine}(2022)}]{penfold2022deep}%
  \BibitemOpen
  \bibfield  {author} {\bibinfo {author} {\bibfnamefont {T.}~\bibnamefont
  {Penfold}}\ and\ \bibinfo {author} {\bibfnamefont {C.}~\bibnamefont
  {Rankine}},\ }\href {https://doi.org/10.1080/00268976.2022.2123406}
  {\bibfield  {journal} {\bibinfo  {journal} {Molecular Physics}\ ,\ \bibinfo
  {pages} {e2123406}} (\bibinfo {year} {2022})}\BibitemShut {NoStop}%
\bibitem [{\citenamefont {Carrasquilla}\ and\ \citenamefont
  {Melko}(2017)}]{Carrasquilla_NP_17_ML_phases}%
  \BibitemOpen
  \bibfield  {author} {\bibinfo {author} {\bibfnamefont {J.}~\bibnamefont
  {Carrasquilla}}\ and\ \bibinfo {author} {\bibfnamefont {R.~G.}\ \bibnamefont
  {Melko}},\ }\href {\doibase 10.1038/nphys4035} {\bibfield  {journal}
  {\bibinfo  {journal} {Nature Physics}\ }\textbf {\bibinfo {volume} {13}},\
  \bibinfo {pages} {431} (\bibinfo {year} {2017})}\BibitemShut {NoStop}%
\bibitem [{\citenamefont {Miles}\ \emph {et~al.}(2021)\citenamefont {Miles},
  \citenamefont {Carbone}, \citenamefont {Sturm}, \citenamefont {Lu},
  \citenamefont {Weichselbaum}, \citenamefont {Barros},\ and\ \citenamefont
  {Konik}}]{Miles_PRB_21_ML_impurity}%
  \BibitemOpen
  \bibfield  {author} {\bibinfo {author} {\bibfnamefont {C.}~\bibnamefont
  {Miles}}, \bibinfo {author} {\bibfnamefont {M.~R.}\ \bibnamefont {Carbone}},
  \bibinfo {author} {\bibfnamefont {E.~J.}\ \bibnamefont {Sturm}}, \bibinfo
  {author} {\bibfnamefont {D.}~\bibnamefont {Lu}}, \bibinfo {author}
  {\bibfnamefont {A.}~\bibnamefont {Weichselbaum}}, \bibinfo {author}
  {\bibfnamefont {K.}~\bibnamefont {Barros}}, \ and\ \bibinfo {author}
  {\bibfnamefont {R.~M.}\ \bibnamefont {Konik}},\ }\href {\doibase
  10.1103/PhysRevB.104.235111} {\bibfield  {journal} {\bibinfo  {journal}
  {Phys. Rev. B}\ }\textbf {\bibinfo {volume} {104}},\ \bibinfo {pages}
  {235111} (\bibinfo {year} {2021})}\BibitemShut {NoStop}%
\bibitem [{\citenamefont {Liu}\ \emph {et~al.}(2017)\citenamefont {Liu},
  \citenamefont {Qi}, \citenamefont {Meng},\ and\ \citenamefont
  {Fu}}]{Liu_PRB_17_ML_MonteCarlo}%
  \BibitemOpen
  \bibfield  {author} {\bibinfo {author} {\bibfnamefont {J.}~\bibnamefont
  {Liu}}, \bibinfo {author} {\bibfnamefont {Y.}~\bibnamefont {Qi}}, \bibinfo
  {author} {\bibfnamefont {Z.~Y.}\ \bibnamefont {Meng}}, \ and\ \bibinfo
  {author} {\bibfnamefont {L.}~\bibnamefont {Fu}},\ }\href {\doibase
  10.1103/PhysRevB.95.041101} {\bibfield  {journal} {\bibinfo  {journal} {Phys.
  Rev. B}\ }\textbf {\bibinfo {volume} {95}},\ \bibinfo {pages} {041101}
  (\bibinfo {year} {2017})}\BibitemShut {NoStop}%
\bibitem [{\citenamefont {Sturm}\ \emph {et~al.}(2021)\citenamefont {Sturm},
  \citenamefont {Carbone}, \citenamefont {Lu}, \citenamefont {Weichselbaum},\
  and\ \citenamefont {Konik}}]{Sturm_PRB_21_ML_impurity}%
  \BibitemOpen
  \bibfield  {author} {\bibinfo {author} {\bibfnamefont {E.~J.}\ \bibnamefont
  {Sturm}}, \bibinfo {author} {\bibfnamefont {M.~R.}\ \bibnamefont {Carbone}},
  \bibinfo {author} {\bibfnamefont {D.}~\bibnamefont {Lu}}, \bibinfo {author}
  {\bibfnamefont {A.}~\bibnamefont {Weichselbaum}}, \ and\ \bibinfo {author}
  {\bibfnamefont {R.~M.}\ \bibnamefont {Konik}},\ }\href {\doibase
  10.1103/PhysRevB.103.245118} {\bibfield  {journal} {\bibinfo  {journal}
  {Phys. Rev. B}\ }\textbf {\bibinfo {volume} {103}},\ \bibinfo {pages}
  {245118} (\bibinfo {year} {2021})}\BibitemShut {NoStop}%
\bibitem [{\citenamefont {Arsenault}\ \emph {et~al.}(2014)\citenamefont
  {Arsenault}, \citenamefont {Lopez-Bezanilla}, \citenamefont {von
  Lilienfeld},\ and\ \citenamefont {Millis}}]{Arsenault_PRB_14_ML_impurity}%
  \BibitemOpen
  \bibfield  {author} {\bibinfo {author} {\bibfnamefont {L.-F. m.~c.}\
  \bibnamefont {Arsenault}}, \bibinfo {author} {\bibfnamefont {A.}~\bibnamefont
  {Lopez-Bezanilla}}, \bibinfo {author} {\bibfnamefont {O.~A.}\ \bibnamefont
  {von Lilienfeld}}, \ and\ \bibinfo {author} {\bibfnamefont {A.~J.}\
  \bibnamefont {Millis}},\ }\href {\doibase 10.1103/PhysRevB.90.155136}
  {\bibfield  {journal} {\bibinfo  {journal} {Phys. Rev. B}\ }\textbf {\bibinfo
  {volume} {90}},\ \bibinfo {pages} {155136} (\bibinfo {year}
  {2014})}\BibitemShut {NoStop}%
\bibitem [{\citenamefont {Walker}\ \emph {et~al.}(2020)\citenamefont {Walker},
  \citenamefont {Kellar}, \citenamefont {Zhang},\ and\ \citenamefont
  {Tam}}]{walker2020neural}%
  \BibitemOpen
  \bibfield  {author} {\bibinfo {author} {\bibfnamefont {N.}~\bibnamefont
  {Walker}}, \bibinfo {author} {\bibfnamefont {S.}~\bibnamefont {Kellar}},
  \bibinfo {author} {\bibfnamefont {Y.}~\bibnamefont {Zhang}}, \ and\ \bibinfo
  {author} {\bibfnamefont {K.-M.}\ \bibnamefont {Tam}},\ }\href@noop {} {\
  (\bibinfo {year} {2020})},\ \Eprint {http://arxiv.org/abs/2008.12331}
  {arXiv:2008.12331} \BibitemShut {NoStop}%
\bibitem [{\citenamefont {Marsiglio}\ \emph {et~al.}(1991)\citenamefont
  {Marsiglio}, \citenamefont {Ruckenstein}, \citenamefont {Schmitt-Rink},\ and\
  \citenamefont {Varma}}]{Marsiglio_PRB_91_SCBA}%
  \BibitemOpen
  \bibfield  {author} {\bibinfo {author} {\bibfnamefont {F.}~\bibnamefont
  {Marsiglio}}, \bibinfo {author} {\bibfnamefont {A.~E.}\ \bibnamefont
  {Ruckenstein}}, \bibinfo {author} {\bibfnamefont {S.}~\bibnamefont
  {Schmitt-Rink}}, \ and\ \bibinfo {author} {\bibfnamefont {C.~M.}\
  \bibnamefont {Varma}},\ }\href {\doibase 10.1103/PhysRevB.43.10882}
  {\bibfield  {journal} {\bibinfo  {journal} {Phys. Rev. B}\ }\textbf {\bibinfo
  {volume} {43}},\ \bibinfo {pages} {10882} (\bibinfo {year}
  {1991})}\BibitemShut {NoStop}%
\bibitem [{\citenamefont {Martinez}\ and\ \citenamefont
  {Horsch}(1991)}]{Martinez_PRB_91_SCBA}%
  \BibitemOpen
  \bibfield  {author} {\bibinfo {author} {\bibfnamefont {G.}~\bibnamefont
  {Martinez}}\ and\ \bibinfo {author} {\bibfnamefont {P.}~\bibnamefont
  {Horsch}},\ }\href {\doibase 10.1103/PhysRevB.44.317} {\bibfield  {journal}
  {\bibinfo  {journal} {Phys. Rev. B}\ }\textbf {\bibinfo {volume} {44}},\
  \bibinfo {pages} {317} (\bibinfo {year} {1991})}\BibitemShut {NoStop}%
\bibitem [{\citenamefont {Liu}\ and\ \citenamefont
  {Manousakis}(1991)}]{Liu_PRB_91_SCBA}%
  \BibitemOpen
  \bibfield  {author} {\bibinfo {author} {\bibfnamefont {Z.}~\bibnamefont
  {Liu}}\ and\ \bibinfo {author} {\bibfnamefont {E.}~\bibnamefont
  {Manousakis}},\ }\href {\doibase 10.1103/PhysRevB.44.2414} {\bibfield
  {journal} {\bibinfo  {journal} {Phys. Rev. B}\ }\textbf {\bibinfo {volume}
  {44}},\ \bibinfo {pages} {2414} (\bibinfo {year} {1991})}\BibitemShut
  {NoStop}%
\bibitem [{\citenamefont {Liu}\ and\ \citenamefont
  {Manousakis}(1992)}]{Liu_PRB_92_SCBA}%
  \BibitemOpen
  \bibfield  {author} {\bibinfo {author} {\bibfnamefont {Z.}~\bibnamefont
  {Liu}}\ and\ \bibinfo {author} {\bibfnamefont {E.}~\bibnamefont
  {Manousakis}},\ }\href {\doibase 10.1103/PhysRevB.45.2425} {\bibfield
  {journal} {\bibinfo  {journal} {Phys. Rev. B}\ }\textbf {\bibinfo {volume}
  {45}},\ \bibinfo {pages} {2425} (\bibinfo {year} {1992})}\BibitemShut
  {NoStop}%
\bibitem [{\citenamefont {Yin}\ and\ \citenamefont
  {Gong}(1997)}]{Yin_PRB_97_SCBA_bilayer}%
  \BibitemOpen
  \bibfield  {author} {\bibinfo {author} {\bibfnamefont {W.-G.}\ \bibnamefont
  {Yin}}\ and\ \bibinfo {author} {\bibfnamefont {C.-D.}\ \bibnamefont {Gong}},\
  }\href {\doibase 10.1103/PhysRevB.56.2843} {\bibfield  {journal} {\bibinfo
  {journal} {Phys. Rev. B}\ }\textbf {\bibinfo {volume} {56}},\ \bibinfo
  {pages} {2843} (\bibinfo {year} {1997})}\BibitemShut {NoStop}%
\bibitem [{\citenamefont {Yin}\ and\ \citenamefont
  {Gong}(1998)}]{Yin_PRB_98_SCBA_multilayer}%
  \BibitemOpen
  \bibfield  {author} {\bibinfo {author} {\bibfnamefont {W.-G.}\ \bibnamefont
  {Yin}}\ and\ \bibinfo {author} {\bibfnamefont {C.-D.}\ \bibnamefont {Gong}},\
  }\href {\doibase 10.1103/PhysRevB.57.11743} {\bibfield  {journal} {\bibinfo
  {journal} {Phys. Rev. B}\ }\textbf {\bibinfo {volume} {57}},\ \bibinfo
  {pages} {11743} (\bibinfo {year} {1998})}\BibitemShut {NoStop}%
\bibitem [{\citenamefont
  {Manousakis}(2007{\natexlab{a}})}]{Manousakis_PRB_07_kink_SCBA}%
  \BibitemOpen
  \bibfield  {author} {\bibinfo {author} {\bibfnamefont {E.}~\bibnamefont
  {Manousakis}},\ }\href {\doibase 10.1103/PhysRevB.75.035106} {\bibfield
  {journal} {\bibinfo  {journal} {Phys. Rev. B}\ }\textbf {\bibinfo {volume}
  {75}},\ \bibinfo {pages} {035106} (\bibinfo {year}
  {2007}{\natexlab{a}})}\BibitemShut {NoStop}%
\bibitem [{\citenamefont
  {Manousakis}(2007{\natexlab{b}})}]{Manousakis_PLA_07_kink_SCBA}%
  \BibitemOpen
  \bibfield  {author} {\bibinfo {author} {\bibfnamefont {E.}~\bibnamefont
  {Manousakis}},\ }\href {\doibase
  https://doi.org/10.1016/j.physleta.2006.11.066} {\bibfield  {journal}
  {\bibinfo  {journal} {Physics Letters A}\ }\textbf {\bibinfo {volume}
  {362}},\ \bibinfo {pages} {86} (\bibinfo {year}
  {2007}{\natexlab{b}})}\BibitemShut {NoStop}%
\bibitem [{\citenamefont {Valla}\ \emph {et~al.}(2007)\citenamefont {Valla},
  \citenamefont {Kidd}, \citenamefont {Yin}, \citenamefont {Gu}, \citenamefont
  {Johnson}, \citenamefont {Pan},\ and\ \citenamefont
  {Fedorov}}]{Valla_PRL_07_kink_SCBA}%
  \BibitemOpen
  \bibfield  {author} {\bibinfo {author} {\bibfnamefont {T.}~\bibnamefont
  {Valla}}, \bibinfo {author} {\bibfnamefont {T.~E.}\ \bibnamefont {Kidd}},
  \bibinfo {author} {\bibfnamefont {W.-G.}\ \bibnamefont {Yin}}, \bibinfo
  {author} {\bibfnamefont {G.~D.}\ \bibnamefont {Gu}}, \bibinfo {author}
  {\bibfnamefont {P.~D.}\ \bibnamefont {Johnson}}, \bibinfo {author}
  {\bibfnamefont {Z.-H.}\ \bibnamefont {Pan}}, \ and\ \bibinfo {author}
  {\bibfnamefont {A.~V.}\ \bibnamefont {Fedorov}},\ }\href {\doibase
  10.1103/PhysRevLett.98.167003} {\bibfield  {journal} {\bibinfo  {journal}
  {Phys. Rev. Lett.}\ }\textbf {\bibinfo {volume} {98}},\ \bibinfo {pages}
  {167003} (\bibinfo {year} {2007})}\BibitemShut {NoStop}%
\bibitem [{\citenamefont {Leung}\ and\ \citenamefont
  {Gooding}(1995)}]{Leung_PRB_95_ED_SCBA}%
  \BibitemOpen
  \bibfield  {author} {\bibinfo {author} {\bibfnamefont {P.~W.}\ \bibnamefont
  {Leung}}\ and\ \bibinfo {author} {\bibfnamefont {R.~J.}\ \bibnamefont
  {Gooding}},\ }\href {\doibase 10.1103/PhysRevB.52.R15711} {\bibfield
  {journal} {\bibinfo  {journal} {Phys. Rev. B}\ }\textbf {\bibinfo {volume}
  {52}},\ \bibinfo {pages} {R15711} (\bibinfo {year} {1995})}\BibitemShut
  {NoStop}%
\bibitem [{\citenamefont {Diamantis}\ and\ \citenamefont
  {Manousakis}(2021)}]{Diamantis_NJP_21_SCBA_MonteCarlo}%
  \BibitemOpen
  \bibfield  {author} {\bibinfo {author} {\bibfnamefont {N.~G.}\ \bibnamefont
  {Diamantis}}\ and\ \bibinfo {author} {\bibfnamefont {E.}~\bibnamefont
  {Manousakis}},\ }\href {\doibase 10.1088/1367-2630/ac39b5} {\bibfield
  {journal} {\bibinfo  {journal} {New Journal of Physics}\ }\textbf {\bibinfo
  {volume} {23}},\ \bibinfo {pages} {123005} (\bibinfo {year}
  {2021})}\BibitemShut {NoStop}%
\bibitem [{\citenamefont {Biau}\ and\ \citenamefont
  {Devroye}(2015)}]{biau_2015}%
  \BibitemOpen
  \bibfield  {author} {\bibinfo {author} {\bibfnamefont {G.}~\bibnamefont
  {Biau}}\ and\ \bibinfo {author} {\bibfnamefont {L.}~\bibnamefont {Devroye}},\
  }\href {\doibase 10.1007/978-3-319-25388-6} {\emph {\bibinfo {title}
  {Lectures on the Nearest Neighbor Method}}}\ (\bibinfo  {publisher} {Springer
  International Publishing},\ \bibinfo {year} {2015})\BibitemShut {NoStop}%
\bibitem [{\citenamefont {Shepard}(1968)}]{Shepard_ACM_68_ML_interpolation}%
  \BibitemOpen
  \bibfield  {author} {\bibinfo {author} {\bibfnamefont {D.}~\bibnamefont
  {Shepard}},\ }\href {\doibase 10.1145/800186.810616} {\bibfield  {journal}
  {\bibinfo  {journal} {Proceedings of the 1968 23rd ACM National Conference}\
  }\bibinfo {series} {ACM '68},\ \bibinfo {pages} {517–524} (\bibinfo {year}
  {1968})}\BibitemShut {NoStop}%
\bibitem [{\citenamefont {Gardner}\ and\ \citenamefont
  {Dorling}(1998)}]{Gardner_review_1998_ML_MLP}%
  \BibitemOpen
  \bibfield  {author} {\bibinfo {author} {\bibfnamefont {M.}~\bibnamefont
  {Gardner}}\ and\ \bibinfo {author} {\bibfnamefont {S.}~\bibnamefont
  {Dorling}},\ }\href {\doibase https://doi.org/10.1016/S1352-2310(97)00447-0}
  {\bibfield  {journal} {\bibinfo  {journal} {Atmospheric Environment}\
  }\textbf {\bibinfo {volume} {32}},\ \bibinfo {pages} {2627} (\bibinfo {year}
  {1998})}\BibitemShut {NoStop}%
\bibitem [{\citenamefont {Kingma}\ and\ \citenamefont
  {Ba}(2014)}]{Kingma_arXiv_14_ML_ADAM}%
  \BibitemOpen
  \bibfield  {author} {\bibinfo {author} {\bibfnamefont {D.~P.}\ \bibnamefont
  {Kingma}}\ and\ \bibinfo {author} {\bibfnamefont {J.}~\bibnamefont {Ba}},\
  }\href {https://arxiv.org/abs/1412.6980} {\bibfield  {journal} {\bibinfo
  {journal} {arXiv:1412.6980}\ } (\bibinfo {year} {2014})}\BibitemShut
  {NoStop}%
\bibitem [{\citenamefont {Wold}\ \emph {et~al.}(1987)\citenamefont {Wold},
  \citenamefont {Esbensen},\ and\ \citenamefont {Geladi}}]{Wold_87_ML_PCA}%
  \BibitemOpen
  \bibfield  {author} {\bibinfo {author} {\bibfnamefont {S.}~\bibnamefont
  {Wold}}, \bibinfo {author} {\bibfnamefont {K.}~\bibnamefont {Esbensen}}, \
  and\ \bibinfo {author} {\bibfnamefont {P.}~\bibnamefont {Geladi}},\ }\href
  {\doibase https://doi.org/10.1016/0169-7439(87)80084-9} {\bibfield  {journal}
  {\bibinfo  {journal} {Chemometrics and Intelligent Laboratory Systems}\
  }\textbf {\bibinfo {volume} {2}},\ \bibinfo {pages} {37} (\bibinfo {year}
  {1987})}\BibitemShut {NoStop}%
\bibitem [{\citenamefont {Lee}\ \emph {et~al.}(2023)\citenamefont {Lee},
  \citenamefont {Carbone},\ and\ \citenamefont
  {Yin}}]{lee_jackson_2023_7527378}%
  \BibitemOpen
  \bibfield  {author} {\bibinfo {author} {\bibfnamefont {J.}~\bibnamefont
  {Lee}}, \bibinfo {author} {\bibfnamefont {M.~R.}\ \bibnamefont {Carbone}}, \
  and\ \bibinfo {author} {\bibfnamefont {W.}~\bibnamefont {Yin}},\ }\href
  {\doibase 10.5281/zenodo.7527378} {\enquote {\bibinfo {title} {{Data
  Repository for: Machine-learning the spectral function of a hole in a quantum
  antiferromagnet}},}\ } (\bibinfo {year} {2023})\BibitemShut {NoStop}%
\bibitem [{\citenamefont {Yin}\ and\ \citenamefont
  {Ku}(2009{\natexlab{b}})}]{Yin_PRB_09_FTBC}%
  \BibitemOpen
  \bibfield  {author} {\bibinfo {author} {\bibfnamefont {W.-G.}\ \bibnamefont
  {Yin}}\ and\ \bibinfo {author} {\bibfnamefont {W.}~\bibnamefont {Ku}},\
  }\href {\doibase 10.1103/PhysRevB.80.180402} {\bibfield  {journal} {\bibinfo
  {journal} {Phys. Rev. B}\ }\textbf {\bibinfo {volume} {80}},\ \bibinfo
  {pages} {180402} (\bibinfo {year} {2009}{\natexlab{b}})}\BibitemShut
  {NoStop}%
\end{thebibliography}
